\documentclass[aps,prd,twocolumn,showpacs,groupedaddress,nofootinbib]{revtex4}

\usepackage{graphicx}  
\usepackage{dcolumn}   
\usepackage{bm}        
\usepackage[english]{babel}
\usepackage{amsfonts,amsmath,amssymb,mathrsfs}
\usepackage{times}
\newcommand{\araa}{Ann.\ Rev.\ Astron.\ Astrophys.}

\newcommand{\cqg}{Class.\ Quantum Grav.}
\newcommand{\mnras}{Mon. Not. Roy. Astron. Soc}
\newcommand{\myprd}{Phys.\ Rev.\ D}
\newcommand{\eq}{\begin{equation}}
\newcommand{\eeq}{\end{equation}}
\newcommand\T{\rule{0pt}{3ex}}
\newcommand\B{\rule[-2ex]{0pt}{0pt}}

\begin{document}

\title{Circular and non-circular nearly horizon-skimming orbits in
  Kerr spacetimes} 
\author{Enrico Barausse}\email{barausse@sissa.it}
\affiliation{SISSA, International School for
             Advanced Studies and INFN, Via Beirut 2, 34014 Trieste, Italy}

\author{Scott A.\ Hughes}
\affiliation{Department of Physics and MIT Kavli Institute, MIT, 77
Massachusetts Ave., Cambridge, MA 02139 USA}

\author{Luciano Rezzolla}
\affiliation{Max-Planck-Institut f\"ur Gravitationsphysik,
             Albert-Einstein-Institut, 14476 Potsdam, Germany}
\affiliation{Department of Physics, Louisiana State University, Baton
             Rouge, LA 70803 USA}

\date{\today}
\begin{abstract}
  We have performed a detailed analysis of orbital motion in the
  vicinity of a nearly extremal Kerr black hole.  For very rapidly
  rotating black holes --- spin parameter $a \equiv J/M > 0.9524M$ ---
  we have found a class of very strong field eccentric orbits whose
  orbital angular momentum $L_z$ increases with the orbit's
  inclination with respect to the equatorial plane, while keeping
  latus rectum and eccentricity fixed.  This behavior is in contrast
  with Newtonian intuition, and is in fact opposite to the ``normal''
  behavior of black hole orbits.  Such behavior was noted previously
  for circular orbits; since it only applies to orbits very close to
  the black hole, they were named ``nearly horizon-skimming orbits''.
  Our current analysis generalizes this result, mapping out the full
  generic (inclined and eccentric) family of nearly horizon-skimming
  orbits.  The earlier work on circular orbits reported that, under
  gravitational radiation emission, nearly horizon-skimming orbits
  exhibit unusual inspiral, tending to evolve to smaller orbit
  inclination, toward prograde equatorial configuration.  Normal
  orbits, by contrast, always demonstrate slowly {\it growing} orbit
  inclination --- orbits evolve toward the retrograde equatorial
  configuration.  Using up-to-date Teukolsky-based fluxes, we have
  concluded that the earlier result was incorrect --- all circular orbits, {\it
  including nearly horizon-skimming ones}, exhibit growing orbit
  inclination under radiative backreaction. Using kludge fluxes based on
  a Post-Newtonian expansion corrected with fits to circular and to equatorial Teukolsky-based fluxes,
  we argue that the inclination grows also for eccentric nearly horizon-skimming orbits.
  We also find that the inclination change is, in any case, very small.  As such, we
  conclude that these orbits are {\it not} likely to have a clear
  and peculiar imprint on the gravitational waveforms expected to be
  measured by the space-based detector LISA.
\end{abstract}

\maketitle

\section{\label{sec:intro}Introduction}

The space-based gravitational-wave detector LISA~\cite{LISA} will be a
unique tool to probe the nature of supermassive black holes (SMBHs),
making it possible to map in detail their spacetimes.  This goal is
expected to be achieved by observing gravitational waves emitted by
compact stars or black holes with masses $\mu \approx 1 -
100\,M_\odot$ spiraling into the SMBHs which reside in the center of
galaxies~\cite{SMBH} (particularly the low end of the galactic center
black hole mass function, $M \approx 10^5 - 10^7\,M_\odot$).  Such
events are known as {\it extreme mass ratio inspirals} (EMRIs).
Current wisdom suggests that several tens to perhaps of order a
thousand such events could be measured per year by
LISA~\cite{gair_event_rates}.

Though the distribution of spins for observed astrophysical black
holes is not very well understood at present, very rapid spin is
certainly plausible, as accretion tends to spin-up
SMBHs~\cite{shapiro}.  Most models for quasi-periodic oscillations (QPOs) suggest this is indeed the case
in all low-mass x-ray binaries for which data is
available~\cite{rezzolla03}.  On the other hand, continuum spectral
fitting of some high-mass x-ray binaries indicates that modest spins
(spin parameter $a/M \equiv J/M^2 \sim 0.6 - 0.8$) are likewise
plausible {\cite{shafee_etal}}.  The continuum-fit technique {\it
does} find an extremely high spin of $a/M \gtrsim 0.98$ for the galactic ``microquasar'' GRS1915+105  {\cite{mcclintock_etal}}.  This argues
for a wide variety of possible spins, depending on the detailed birth
and growth history of a given black hole.

In the mass range corresponding to black holes in galactic centers,
measurements of the broad iron K$\alpha$ emission line in active
galactic nuclei suggest that SMBHs can be very rapidly rotating (see
Ref.\ \cite{iron_line_rev} for a recent review).  For instance, in the
case of MCG-6-30-15, for which highly accurate observations are
available, $a$ has been found to be larger than $0.987M$ at $90\%$
confidence~\cite{iron_line_res}.  Because gravitational waves from
EMRIs are expected to yield a very precise determination of the spins
of SMBHs~\cite{spin_measurements}, it is interesting to investigate
whether EMRIs around very rapidly rotating black holes may possess
peculiar features which would be observable by LISA. Should such
features exist, they would provide unambiguous information on the spin
of SMBHs and thus on the mechanisms leading to their
formation~\cite{volonteri}.

For extremal Kerr black holes ($a = M$), the existence of a special
class of ``circular'' orbits was pointed out long ago by
Wilkins~\cite{wilkins}, who named them ``horizon-skimming'' orbits.
(``Circular'' here means that the orbits are of constant
Boyer-Lindquist coordinate radius $r$.)  These orbits have varying
inclination angle with respect to the equatorial plane and have the
same coordinate radius as the horizon, $r = M$.  Despite this
seemingly hazardous location, it can be shown that all these $r = M$
orbits have finite separation from one another and from the event
horizon {\cite{bpt}}.  Their somewhat pathological description is due
to a singularity in the Boyer-Lindquist coordinates, which collapses a
finite span of the spacetime into $r = M$.

Besides being circular and ``horizon-skimming'', these orbits also
show peculiar behavior in their relation of angular momentum to
inclination.  In Newtonian gravity, a generic orbit has
$L_z=|\mbox{\boldmath$L$}|\cos\iota$, where $\iota$ is the inclination
angle relative to the equatorial plane (going from $\iota = 0$ for
equatorial prograde orbits to $\iota = \pi$ for equatorial retrograde
orbits, passing through $\iota = \pi/2$ for polar orbits), and
$\mbox{\boldmath$L$}$ is the orbital angular momentum vector.  As a
result, $\partial L_z(r, \iota)/\partial\iota < 0$, and the angular
momentum in the $z$-direction always decreases with increasing
inclination if the orbit's radius is kept constant.  This intuitively
reasonable decrease of $L_z$ with $\iota$ is seen for almost all black
hole orbits as well.  Horizon-skimming orbits, by contrast, exhibit
exactly the opposite behavior: $L_z$ increases with inclination angle.

Reference \cite{skimming} asked whether the behavior $\partial
L_z/\partial\iota > 0$ could be extended to a broader class of
circular orbits than just those at the radius $r = M$ for the spin
value $a = M$.  It was found that this condition is indeed more
general, and extended over a range of radius from the ``innermost
stable circular orbit'' to $r \simeq 1.8M$ for black holes with $a >
0.9524M$.  Orbits that show this property have been named ``nearly
horizon-skimming''.  The Newtonian behavior $\partial L_z(r,
\iota)/\partial \iota < 0$ is recovered for all orbits at $r\gtrsim
1.8M$~\cite{skimming}.

A qualitative understanding of this behavior comes from recalling that
very close to the black hole all physical processes become ``locked''
to the hole's event horizon~\cite{membrane}, with the orbital motion
of point particles coupling to the horizon's spin. This locking
dominates the ``Keplerian'' tendency of an orbit to move more quickly
at smaller radii, forcing an orbiting particle to slow down in the
innermost orbits.  Locking is particularly strong for the most-bound
(equatorial) orbits; the least-bound orbits (which have the largest
inclination) do not strongly lock to the black hole's spin until they
have very nearly reached the innermost orbit~\cite{skimming}.  The
property $\partial L_z(r, \iota)/\partial\iota > 0$ reflects the
different efficiency of nearly horizon-skimming orbits to lock with
the horizon.

Reference \cite{skimming} argued that this behavior could have
observational consequences.  It is well-known that the inclination
angle of an inspiraling body generally increases due to
gravitational-wave emission~\cite{fintan_circ,first_code}. Since
$dL_z/dt < 0$ because of the positive angular momentum carried away by
the gravitational waves, and since ``normal'' orbits have $\partial
L_z/\partial \iota < 0$, one would expect $d\iota/dt > 0$.  However,
if during an evolution $\partial L_z/\partial \iota$ switches sign,
then $d\iota/dt$ might switch sign as well: An inspiraling body could
evolve towards an equatorial orbit, signalling the presence of an
``almost-extremal'' Kerr black hole.

It should be emphasized that this argument is not rigorous.  In
particular, one needs to consider the joint evolution of orbital
radius and inclination angle; and, one must include the dependence of
these two quantities on orbital energy as well as angular
momentum\footnote{In the general case, one must also include the
dependence on ``Carter's constant'' $Q$~\cite{Carter}, the third integral of black
hole orbits (described more carefully in Sec.\
\ref{sec:constants_of_motion}).  For circular orbits, $Q = Q(E,L_z)$:
knowledge of $E$ and $L_z$ completely determines $Q$.}.  As such,
$d\iota/dt$ depends not only on $dL_z/dt$ and $\partial L_z/\partial
\iota$, but also on $dE/dt$, $\partial E/\partial\iota$, $\partial
E/\partial r$ and $\partial L_z/\partial r$.

In this sense, the argument made in Ref.\ \cite{skimming} should be
seen as claiming that the contribution coming from $dL_z/dt$ and
$\partial L_z/\partial \iota$ are simply the dominant ones. Using the
numerical code described in~\cite{first_code} to compute the fluxes
$dL_z/dt$ and $dE/dt$, it was then found that a test-particle on a
circular orbit passing through the nearly horizon-skimming region of a
Kerr black hole with $a = 0.998M$ (the value at which a hole's spin
tends to be buffered due to photon capture from thin disk
accretion~\cite{thorne_spin}) had its inclination angle decreased by
$\delta\iota\approx1^\circ - 2^\circ$~\cite{skimming} in the adiabatic approximation~\cite{Mino_adiabatic}.  It should be
noted at this point that the rate of change of inclination angle,
$d\iota/dt$, appears as the {\it difference} of two relatively small
and expensive to compute rates of change [cf.\ Eq.\ (3.8) of Ref.\
{\cite{first_code}}].  As such, small relative errors in those rates
of change can lead to large relative errors in $d\iota/dt$.  Finally,
in Ref.\ {\cite{skimming}} it was speculated that the decrease could
be even larger for eccentric orbits satisfying the condition $\partial
L_z/\partial \iota > 0$, possibly leading to an observable imprint on
EMRI gravitational waveforms.

The main purpose of this paper is to extend Ref.\ \cite{skimming}'s
analysis of nearly horizon-skimming orbits to include the effect of
orbital eccentricity, and to thereby test the speculation that there
may be an observable imprint on EMRI waveforms of nearly
horizon-skimming behavior.  In doing so, we have revisited all the
calculations of Ref.\ {\cite{skimming}} using a more accurate
Teukolsky solver 
which serves as the engine for the analysis
presented in Ref.\ {\cite{hughes_drasco}}.

We have found that the critical spin value for circular nearly
horizon-skimming orbits, $a > 0.9524M$, also delineates a family of
eccentric orbits for which the condition $\partial
L_z(p,e,\iota)/\partial \iota > 0$ holds.  (More precisely, we
consider variation with respect to an angle $\theta_{\rm inc}$ that is
easier to work with in the extreme strong field, but that is easily
related to $\iota$.)  The parameters $p$ and $e$ are the orbit's latus
rectum and eccentricity, defined precisely in Sec.\
{\ref{sec:constants_of_motion}}.  These generic nearly
horizon-skimming orbits all have $p \lesssim 2M$, deep in the black
hole's extreme strong field.

We next study the evolution of these orbits under gravitational-wave
emission in the adiabatic approximation.  We first revisited the evolution of circular, nearly
horizon-skimming orbits using the improved Teukolsky solver which was
used for the analysis of Ref.\ {\cite{hughes_drasco}}.  The results of
this analysis were somewhat surprising: Just as for ``normal'' orbits,
we found that orbital inclination {\it always} increases during
inspiral, even in the nearly horizon-skimming regime.  This is in
stark contrast to the claims of Ref.\ {\cite{skimming}}.  As noted
above, the inclination's rate of change depends on the difference of
two expensive and difficult to compute numbers, and thus can be
strongly impacted by small relative errors in those numbers.  A
primary result of this paper is thus to retract the claim of Ref.\
{\cite{skimming}} that an important dynamic signature of the nearly
horizon-skimming region is a reversal in the sign of inclination angle
evolution: The inclination {\it always} grows under gravitational
radiation emission.

We next extended this analysis to study the evolution of generic
nearly horizon-skimming orbits.  The Teukolsky code to which we have
direct access can, at this point, only compute the radiated fluxes of
energy $E$ and angular momentum $L_z$; results for the evolution of
the Carter constant $Q$ are just now beginning to be understood
{\cite{sago}}, and have not yet been incorporated into this code.  We
instead use ``kludge'' expressions for $dE/dt$, $dL_z/dt$, and $dQ/dt$
which were inspired by Refs.\ {\cite{kludge_hughes,
GG_kludge_fluxes}}.  These expression are based on post-Newtonian flux
formulas, modified in such a way that they fit strong-field radiation
reaction results obtained from a Teukolsky integrator; see Ref.\
{\cite{GG_kludge_fluxes}} for further discussion.  Our analysis
indicates that, just as in the circular limit, the result $d\iota/dt >
0$ holds for generic nearly horizon-skimming orbits.  Furthermore, and
contrary to the speculation of Ref.\ {\cite{skimming}}, we do {\it
not} find a large amplification of $d\iota/dt$ as orbits are made more
eccentric.

Our conclusion is that the nearly horizon-skimming regime, though an
interesting curiosity of strong-field orbits of nearly extremal black
holes, will {\it not} imprint any peculiar observational signature on
EMRI waveforms.

The remainder of this paper is organized as follows. In Sec.\ II, we
review the properties of bound stable orbits in Kerr, providing
expressions for the constants of motion which we will use in Sec.\ III
to generalize nearly horizon-skimming orbits to the non-circular
case. In Sec.\ IV, we study the evolution of the inclination angle for
circular nearly horizon-skimming orbits using Teukolsky-based fluxes;
in Sec.\ V we do the same for non-circular orbits and using kludge
fluxes.  We present and discuss our detailed conclusions in Sec.\ VI.
The fits and post-Newtonian fluxes used for the kludge fluxes are
presented in the Appendix. Throughout the paper we have used units in
which $G = c = 1$.

\section{\label{sec:constants_of_motion} Bound stable orbits in Kerr
  spacetimes} 

\begin{figure*}
\includegraphics[width=8.5cm]{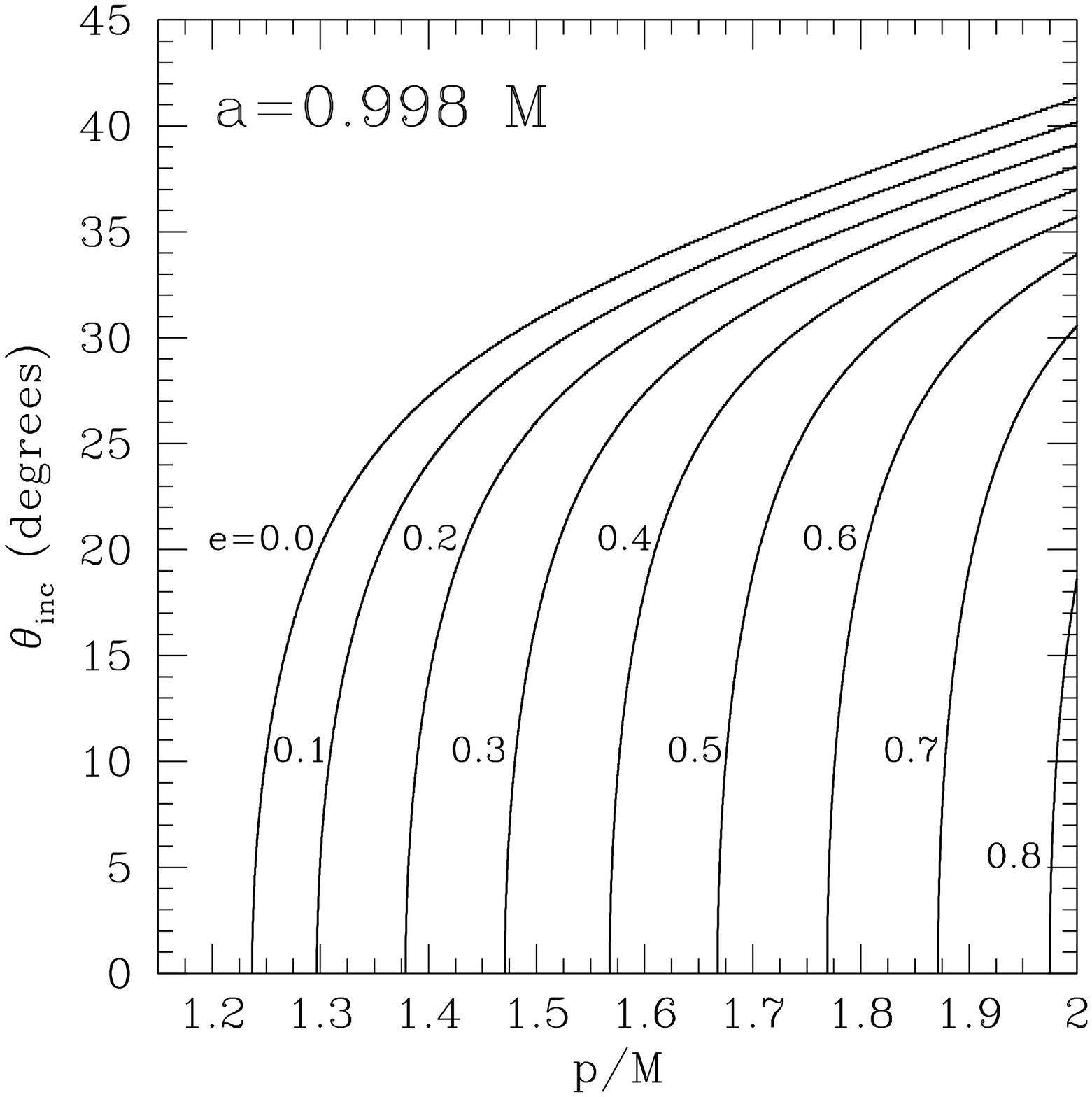}
\hskip 0.5 cm
\includegraphics[width=8.5cm]{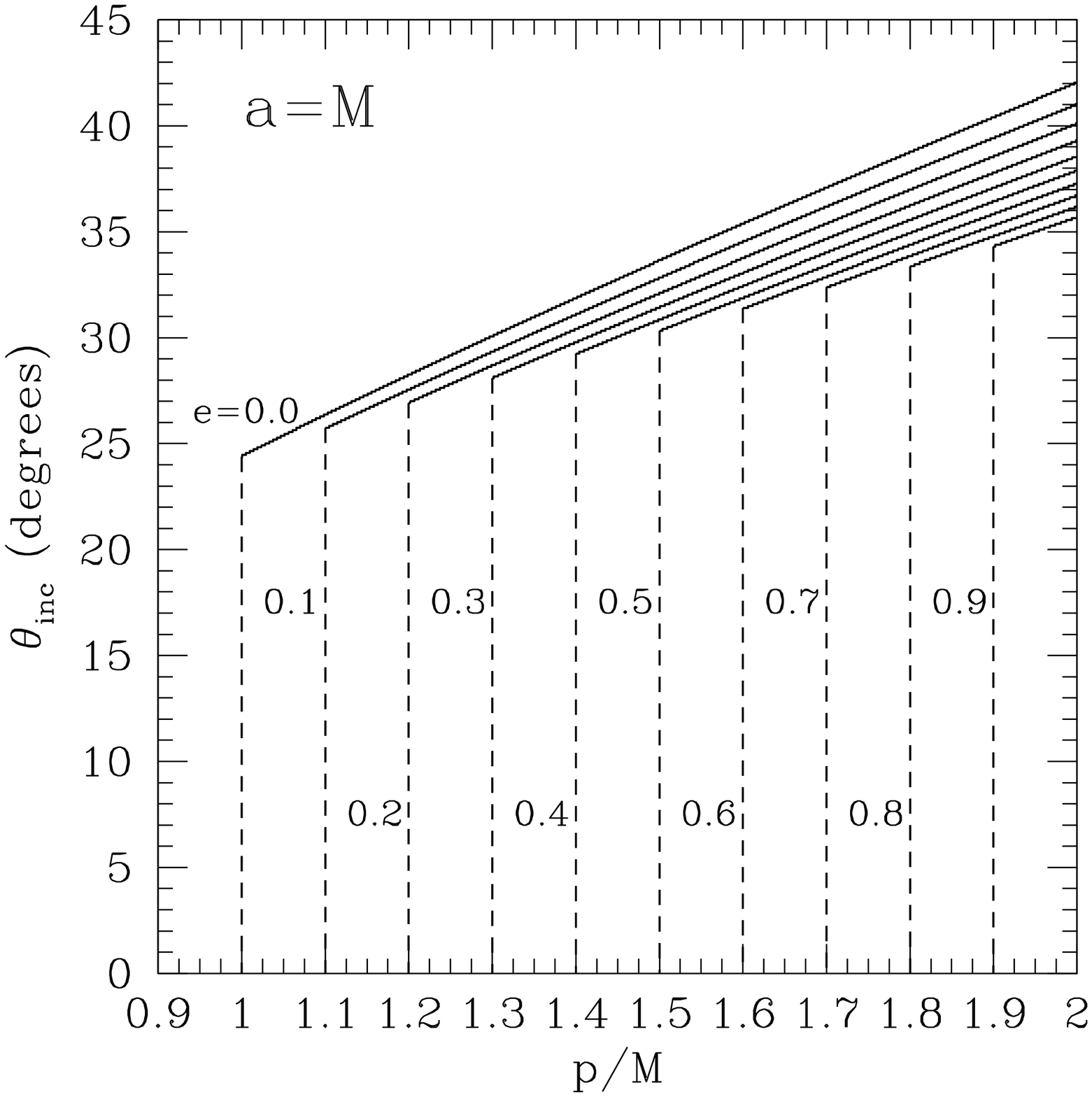}
\caption{\textit{Left panel}: 
Inclination angles $\theta_{\rm inc}$ for which bound stable
orbits exist for a black hole with spin $a = 0.998\,M$. The allowed
range for $\theta_{\rm inc}$ goes from $\theta_{\rm inc}=0$ to the
curve corresponding to the eccentricity under consideration,
$\theta_{\rm inc}=\theta_{\rm inc}^{\rm max}$.
\textit{Right panel}:
Same as left but for an extremal black
hole, $a=M$. Note that in this case $\theta_{\rm inc}^{\rm max}$
never reaches zero.}
\label{fig:allowed}
\end{figure*}

The line element of a Kerr spacetime, written in Boyer-Lindquist
coordinates reads~\cite{MTW}
\begin{eqnarray} 
\label{Kerr}
ds^2 &=& 
- \left( 1-\frac{2Mr}{\Sigma} \right) ~dt^2
+ \frac{\Sigma}{\Delta}~dr^2
+ \Sigma~d\theta^2 \nonumber \\
&&+ \left( r^2+a^2 + \frac{2Ma^2r}{\Sigma}\sin^2\theta \right)\sin^2\theta~d\phi^2 \nonumber \\
&&- \frac{4Mar}{\Sigma}\sin^2\theta~dt~d\phi,
\end{eqnarray}
where 
\begin{align}
& \Sigma \equiv r^2 + a^2\cos^2\theta, & \Delta \equiv r^2 - 2Mr + a^2. &
\end{align}
Up to initial conditions, geodesics can then be labelled by four
constants of motion: the mass $\mu$ of the test particle, its energy
$E$ and angular momentum $L_z$ as measured by an observer at infinity
and the Carter constant $Q$ \cite{Carter}. The presence of these four
conserved quantities makes the geodesic equations separable in
Boyer-Lindquist coordinates. Introducing the Carter time $\lambda$,
defined by
\eq
\frac{d\tau}{d\lambda}\equiv\Sigma\;,\eeq 
the geodesic equations become
\begin{align}
&\left(\mu\frac{dr}{d\lambda}\right)^2 = V_r(r), &
&\mu\frac{dt}{d\lambda} = V_t(r,\theta),&
\nonumber \\
&\left(\mu\frac{d\theta}{d\lambda}\right)^2 = V_\theta(\theta), &
&\mu\frac{d\phi}{d\lambda} = V_\phi(r,\theta)\;,&
 \label{geodesics}
\end{align}
with 
\begin{subequations} \label{detailed geodesics}
\begin{align}
&V_t(r,\theta)
  \equiv E \left( \frac{\varpi^4}{\Delta} - a^2\sin^2\theta \right)
     + aL_z \left( 1 - \frac{\varpi^2}{\Delta} \right),&
\label{tdot} \\
&V_r(r)
  \equiv \left( E\varpi^2 - a L_z \right)^2
  - \Delta\left[\mu^2 r^2 + (L_z - a E)^2 + Q\right],&
\label{rdot}\\
&V_\theta(\theta) \equiv Q - L_z^2 \cot^2\theta - a^2(\mu^2 - E^2)\cos^2\theta,&
\label{thetadot}\\
&V_\phi(r,\theta)
  \equiv L_z \csc^2\theta + aE\left(\frac{\varpi^2}{\Delta} - 1\right) - \frac{a^2L_z}{\Delta},&
\label{phidot}
\end{align}
\end{subequations}
where we have defined
\begin{equation}
\varpi^2 \equiv r^2 + a^2\;.
\end{equation}

The conserved parameters $E$, $L_z$, and $Q$ can be remapped to other
parameters that describe the geometry of the orbit.  We have found it
useful to describe the orbit in terms of an angle $\theta_{\rm min}$
--- the minimum polar angle reached by the orbit --- as well as the
latus rectum $p$ and the eccentricity $e$.  In the weak-field limit,
$p$ and $e$ correspond exactly to the latus rectum and eccentricity
used to describe orbits in Newtonian gravity; in the strong field,
they are essentially just a convenient remapping of the orbit's
apoastron and periastron:
\begin{equation}
r_{\rm ap}\equiv\frac{p}{1-e}\;,\quad r_{\rm
peri}\equiv\frac{p}{1+e}\;.
\end{equation}
Finally, in much of our analysis, it is useful to refer to
\begin{gather}
z_{-}\equiv\cos^2\theta_{\min}\;,
\end{gather}
rather than to $\theta_{\rm min}$ directly.

To map $(E,L_z,Q)$ to $(p,e,z_{-})$, use Eq.\ (\ref{geodesics}) to
impose ${dr}/{d\lambda}=0$ at $r=r_{\rm ap}$ and $r=r_{\rm peri}$, and
to impose ${d\theta}/{d\lambda}=0$ at $\theta = \theta_{\rm min}$.
(Note that for a circular orbit, $r_{\rm ap} = r_{\rm peri} = r_0$.
In this case, one must apply the rules ${dr}/{d\lambda}=0$ and
${d^2r}/{d\lambda^2}=0$ at $r=r_0$.)  Following this approach,
Schmidt~\cite{schmidt} was able to derive explicit expressions for
$E$, $L_z$ and $Q$ in terms of $p$, $e$ and $z_{-}$.  We now briefly
review Schmidt's results.
 
Let us first introduce the dimensionless quantities
\begin{gather}
\tilde{E}\equiv E/\mu\;,\quad
\tilde{L}_z\equiv L_z/(\mu M)\;,\quad\tilde{Q}\equiv Q/(\mu M)^2\;,\\ 
\tilde{a}\equiv a/M\;,\quad
\tilde{r}\equiv r/M\;,\quad \tilde{\Delta}\equiv\Delta/M^2\;,
\end{gather}
and the functions
\begin{eqnarray}
&&  f(\tilde{r}) \equiv \tilde{r}^{4} + \tilde{a}^{2}\left[\tilde{r}(\tilde{r}+2)+z_{-}\tilde{\Delta}\right]\;, \\
&&  g(\tilde{r}) \equiv 2\,\tilde{a}\,\tilde{r}\;, \\
&&  h(\tilde{r}) \equiv \tilde{r}(\tilde{r}-2) + \frac{z_{-}}{1-z_{-}}\tilde{\Delta}\;, \\
&&  d(\tilde{r}) \equiv (\tilde{r}^{2} + \tilde{a}^{2}z_{-})\tilde{\Delta}\;.
\end{eqnarray}
Let us further define the set of functions
\begin{multline}
  (f_{1}, g_{1}, h_{1}, d_{1})  \equiv\\
  \left\{\begin{array}{ll} 
    (f(\tilde{r}_{\mathrm{p}}), g(\tilde{r}_{\mathrm{p}}), 
     h(\tilde{r}_{\mathrm{p}}), d(\tilde{r}_{\mathrm{p}})) &
    \mbox{if $e>0$}\;, \\
    (f(\tilde{r}_{0}), g(\tilde{r}_{0}), h(\tilde{r}_{0}), d(\tilde{r}_{0})) &
    \mbox{if $e=0$}\;,
  \end{array}\right.
\end{multline}
\begin{multline}
  (f_{2}, g_{2}, h_{2}, d_{2}) \equiv\\
  \left\{\begin{array}{ll}
    (f(\tilde{r}_{\mathrm{a}}), g(\tilde{r}_{\mathrm{a}}), 
     h(\tilde{r}_{\mathrm{a}}), d(\tilde{r}_{\mathrm{a}})) & 
    \mbox{if $e>0$}\;, \\
    (f'(\tilde{r}_{0}), g'(\tilde{r}_{0}), h'(\tilde{r}_{0}),
     d'(\tilde{r}_{0})) & 
    \mbox{if $e=0$}\;,
  \end{array}\right.
\end{multline}
and the determinants
\begin{eqnarray}
\kappa &\equiv& d_1h_2 - d_2h_1\;,\label{kappa_def}
\\
\varepsilon &\equiv& d_1g_2 - d_2g_1\;,
\\
\rho &\equiv& f_1h_2 - f_2h_1\;,
\\
\eta &\equiv& f_1g_2 - f_2g_1\;,
\\
\sigma &\equiv& g_1h_2 - g_2h_1\;.
\end{eqnarray}

The energy of the particle can then be written
\begin{equation}
  \tilde{E} =
  \sqrt{\frac{\kappa\rho+2\epsilon\sigma - 
        2D\sqrt{\sigma(\sigma\epsilon^{2}+\rho\epsilon\kappa-\eta\kappa^{2})}
        }{\rho^{2}+4\eta\sigma}}\;.\label{energy_def}
\end{equation}
The parameter $D$ takes the values $\pm1$.  The angular momentum is
a solution of the system
\begin{eqnarray}
      f_{1}\tilde{E}^{2} - 2 g_{1}\tilde{E}\tilde{L}_{z} - 
      h_{1}\tilde{L}_{z}^{2} - d_{1} & = 0\;,
  \label{eq:Lzsys1}\\
      f_{2}\tilde{E}^{2} - 2 g_{2}\tilde{E}\tilde{L}_{z} - 
      h_{2}\tilde{L}_{z}^{2} - d_{2} & = 0\;.
  \label{eq:Lzsys2}
 \end{eqnarray}
By eliminating the $\tilde{L}_z^2$ terms in these equations, one finds
the solution
\eq\label{L_no_sqrt}
\tilde{L}_z = \frac{\rho\tilde{E}^2-\kappa}{2\tilde{E}\sigma}
\eeq
for the angular momentum.  Using $d\theta/d\lambda = 0$ at $\theta =
\theta_{\rm min}$, the Carter constant can be written
\begin{equation}
  \tilde{Q} = z_{-}\left[\tilde{a}^2(1-\tilde{E}^2) +
  \frac{\tilde{L}_z^2}{1-z_{-}}\right]\;.
\end{equation}

Additional constraints on $p$, $e$, $z_{-}$ are needed for the orbits
to be stable. Inspection of Eq.\ (\ref{geodesics}) shows that an
eccentric orbit is stable only if
\eq \frac{\partial V_r}{\partial
  r}(r_{\rm peri}) >0\;. \eeq 
It is marginally stable if ${\partial V_r}/{\partial r}=0$ at $r =
r_{\rm peri}$. Similarly, the stability condition for circular orbits
is
\eq \frac{\partial^2 V_r}{\partial r^2}(r_0) <0\;;
\eeq
marginally stable orbits are set by ${\partial^2 V_r}/{\partial
r^2}=0$ at $r=r_0$.

Finally, we note that one can massage the above solutions for the
conserved orbital quantities of bound stable orbits to rewrite the
solution for $\tilde{L}_z$ as
\eq
\tilde{L}_z = -\frac{g_1 \tilde{E}}{h_1} + \frac{D}{h_1} \sqrt{g_1^2
\tilde{E}^2 + (f_1 \tilde{E}^2 - d_1)h_1}\;.
\label{ang_mom2}
\eeq
From this solution, we see that it is quite natural to refer to orbits
with $D=1$ as \textit{prograde} and to orbits with $D=-1$ as
\textit{retrograde}. Note also that Eq.\ (\ref{ang_mom2}) is a more
useful form than the corresponding expression, Eq.\ (A4), of Ref.\
\cite{hughes_drasco}.  In that expression, the factor $1/h_1$ has been
squared and moved inside the square root.  This obscures the fact that
$h_1$ changes sign for very strong field orbits.  Differences between
Eq.\ (\ref{ang_mom2}) and Eq.\ (A4) of {\cite{hughes_drasco}} are
apparent for $a\gtrsim0.835$, although only for orbits close to the
separatrix (\textit{i.e.}, the surface in the space of parameters
$(p,e,\iota)$ where marginally stable bound orbits lie).

\begin{figure*}
\includegraphics[width=8.5cm]{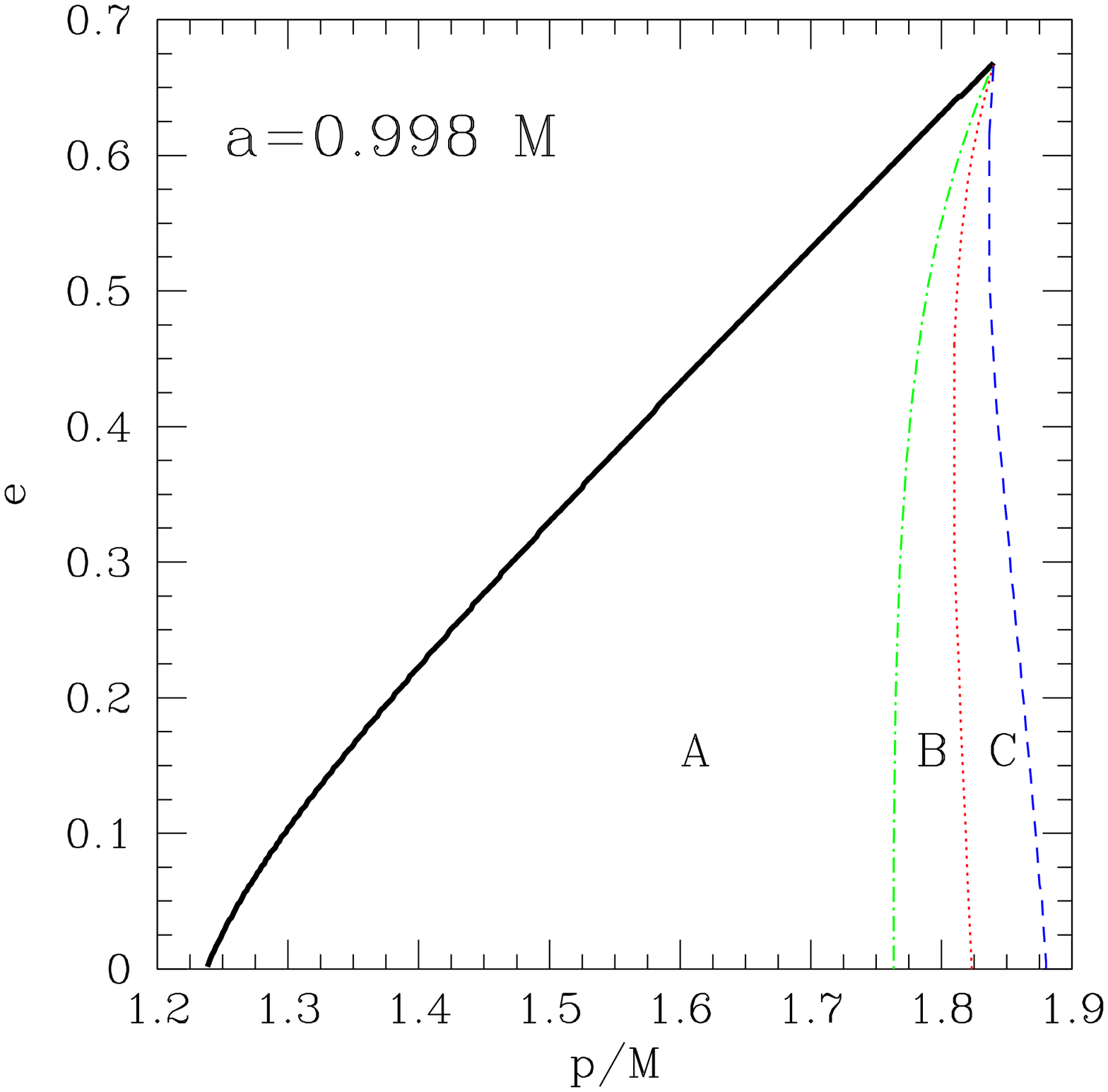}
\hskip 0.5 cm
\includegraphics[width=8.5cm]{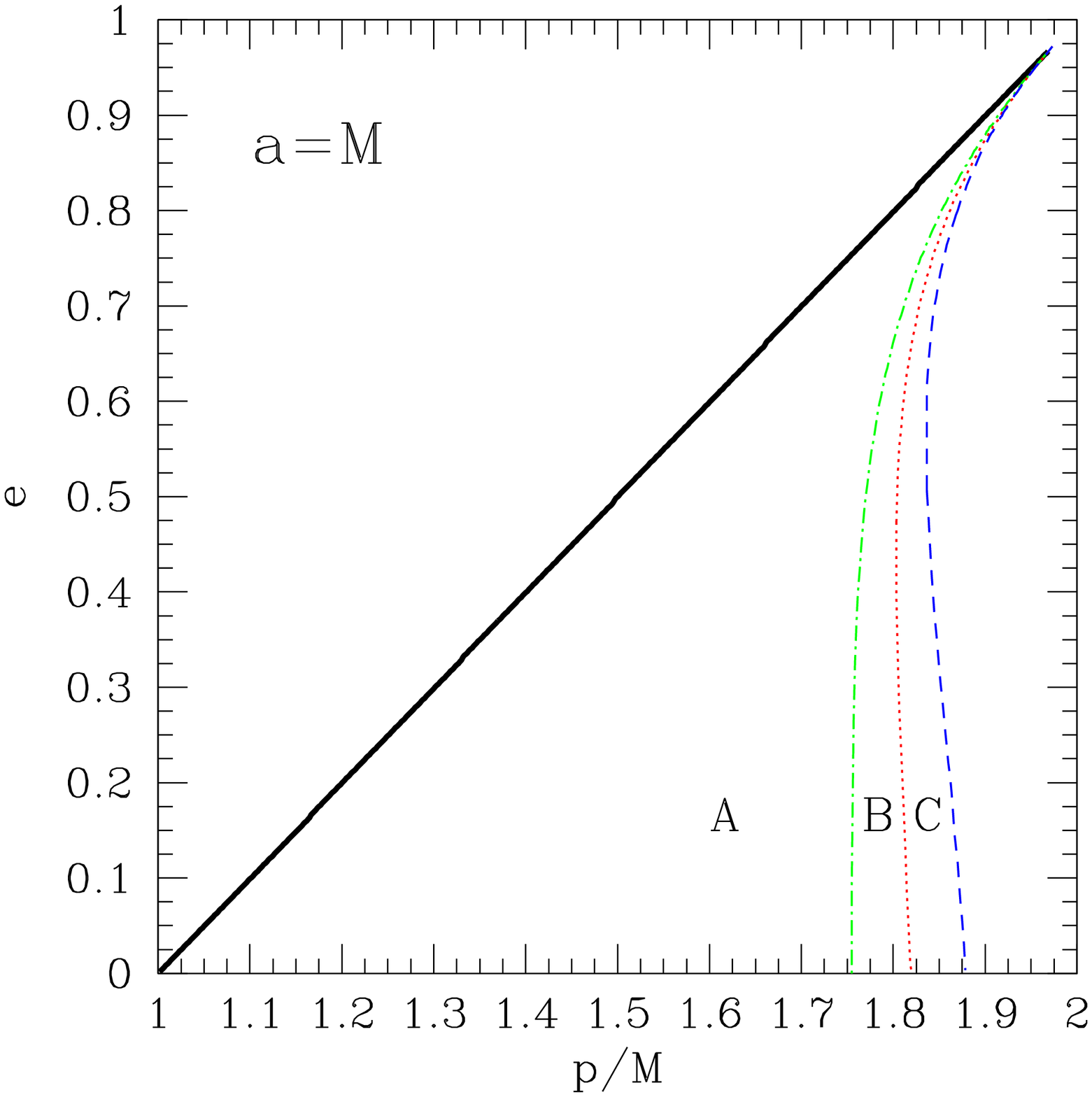}
\caption{\textit{Left panel}: Non-circular nearly horizon-skimming
  orbits for $a = 0.998M$.  The heavy solid line indicates the
  separatrix between stable and unstable orbits for equatorial orbits
  ($\iota = \theta_{\rm inc} = 0$).  All orbits above and to the left
  of this line are unstable.  The dot-dashed line (green in the color
  version) bounds the region of the $(p,e)$-plane where $\partial
  L_z/\partial\theta_{\rm inc} > 0$ for all allowed inclination angles
  (``Region \textit{A}'').  All orbits between this line and the
  separatrix belong to Region \textit{A}.  The dotted line (red in the
  color version) bounds the region $(L_z)_{\rm most\,bound}<(L_z)_{\rm
  least\,bound}$ (``Region \textit{B}'').  Note that \textit{B}
  includes \textit{A}.  The dashed line (blue in the color version)
  bounds the region where ${\partial L_z}/{\partial \theta_{\rm
  inc}}>0$ for at least one inclination angle (``Region \textit{C}'');
  note that \textit{C} includes \textit{B}.  All three of these
  regions are candidate generalizations of the notion of nearly
  horizon-skimming orbits.  \textit{Right panel}: Same as the left
  panel, but for the extreme spin case, $a = M$.  In this case the
  separatrix between stable and unstable equatorial orbits is given by
  the line $p/M=1+e$.}
\label{fig:regions}
\end{figure*}

\section{\label{sec:skimming} Non-circular nearly horizon-skimming orbits}

With explicit expressions for $E$, $L_z$ and $Q$ as functions of $p$,
$e$ and $z_{-}$, we now examine how to generalize the condition
$\partial L_z(r,\iota)/\partial\iota>0$, which defined circular nearly
horizon-skimming orbits in Ref.~\cite{skimming}, to encompass the
non-circular case. We recall that the inclination angle $\iota$ is
defined as~\cite{skimming}
\eq
\label{iota_def}
\cos\iota=\frac{L_z}{\sqrt{Q+{L^2_z}}}\;.
\eeq
Such a definition is not always easy to handle in the case of
eccentric orbits. In addition, $\iota$ does not have an obvious
physical interpretation (even in the circular limit), but rather was
introduced essentially to generalize (at least formally) the
definition of inclination for Schwarzschild black hole orbits.  In
that case, one has $Q={L^2_x}+{L^2_y}$ and therefore $L_z =
|\mbox{\boldmath$L$}|\cos\iota$.

A more useful definition for the inclination angle in Kerr was
introduced in Ref.~\cite{hughes_drasco}:
\eq
\theta_{\rm inc} = \frac{\pi}{2}-D\,\theta_{\min}\;,
\eeq
where $\theta_{\min}$ is the minimum reached by $\theta$ during the
orbital motion.  This angle is trivially related to $z_{-}$
($z_{-}=\sin^2\theta_{\rm inc}$) and ranges from $0$ to $\pi/2$ for
prograde orbits and from $\pi/2$ to $\pi$ for retrograde orbits.  It
is a simple numerical calculation to convert between $\iota$ and
$\theta_{\rm inc}$; doing so shows that the differences between
$\iota$ and $\theta_{\rm inc}$ are very small, with the two coinciding
for $a=0$, and with a difference that is less than $2.6^\circ$ for
$a=M$ and circular orbits with $r=M$.

Bearing all this in mind, the condition we have adopted to generalize
nearly horizon-skimming orbits is
\eq
\frac{\partial L_z(p,e,\theta_{\rm inc})}{\partial\theta_{\rm inc}}>0\;.
\eeq
We have found that certain parts of this calculation, particular the
analysis of strong-field geodesic orbits, are best done using the
angle $\theta_{\rm inc}$; other parts are more simply done using the
angle $\iota$, particularly the ``kludge'' computation of fluxes
described in Sec.\ {\ref{sec:iota_evolution_eccentric}}.  (This is
because the kludge fluxes are based on an extension of post-Newtonian
formulas to the strong-field regime, and these formulas use $\iota$
for inclination angle.)  Accordingly, we often switch back and forth
between these two notions of inclination, and in fact present our
final results for inclination evolution using both $d\iota/dt$ and
$d\theta_{\rm inc}/dt$.

Before mapping out the region corresponding to nearly horizon-skimming
orbits, it is useful to examine stable orbits more generally in the
strong field of rapidly rotating black holes.  We first fix a value
for $a$, and then discretize the space of parameters $(p,e,\theta_{\rm
inc})$.  We next identify the points in this space corresponding to
bound stable geodesic orbits.  Sufficiently close to the horizon, the
bound stable orbits with specified values of $p$ and $e$ have an
inclination angle $\theta_{\rm inc}$ ranging from $0$ (equatorial
orbit) to a maximum value $\theta_{\rm inc}^{\rm max}$.  For given $p$
and $e$, $\theta_{\rm inc}^{\rm max}$ defines the {\it separatrix}
between stable and unstable orbits.

Example separatrices are shown in Fig.~\ref{fig:allowed}
for $a=0.998M$ and $a=M$.  This
figure shows the behavior of $\theta_{\rm inc}^{\rm max}$ as a
function of the latus rectum for the different values of the
eccentricity indicated by the labels.  Note that for $a=0.998M$ the
angle $\theta_{\rm inc}^{\rm max}$ eventually goes to zero.  This is
the general behavior for $a<M$. On the other hand, for an extremal
black hole, $a = M$, $\theta_{\rm inc}^{\rm max}$ never goes to
zero.  The orbits which reside at $r = M$ (the circular limit) are the
``horizon-skimming orbits'' identified by Wilkins {\cite{wilkins}};
the $a = M$ separatrix has a similar shape even for eccentric orbits.
As expected, we find that for given latus rectum and eccentricity the
orbit with $\theta_{\rm inc}=0$ is the one with the lowest energy $E$
(and hence is the most-bound orbit), whereas the orbit with
$\theta_{\rm inc}=\theta_{\rm inc}^{\rm max}$ has the highest $E$
(and is least bound).

Having mapped out stable orbits in $(p,e,\theta_{\rm inc})$ space, we
then computed the partial derivative $\partial L_z(p,e,\theta_{\rm
inc})/\partial\theta_{\rm inc}$ and identified the following three
overlapping regions:

\begin{itemize}

\item \textit{Region A}: The portion of the $(p,e)$ plane for which
  $\partial L_z(p,e,\theta_{\rm inc})/\partial\theta_{\rm inc}>0$ for
  $0 \le \theta_{\rm inc} \le \theta_{\rm inc}^{\rm max}$.  This
  region is illustrated in Fig.\ {\ref{fig:regions}} as the area under
  the heavy solid line and to the left of the dot-dashed line (green in
  the color version).

\item \textit{Region B}: The portion of the $(p,e)$ plane for which
  $(L_z)_{\rm most\,bound}(p,e)$ is smaller than $(L_z)_{\rm
  least\,bound}(p,e)$.  In other words,
\begin{equation}
L_z(p,e,0) < L_z(p,e,\theta_{\rm inc}^{\rm max})
\end{equation}
  in Region \textit{B}.  Note that Region \textit{B} contains Region
  \textit{A}.  It is illustrated in Fig.\ {\ref{fig:regions}} as the
  area under the heavy solid line and to the left of the dotted line
  (red in the color version).

\item \textit{Region C}: The portion of the $(p,e)$ plane for which
  $\partial L_z(p,e,\theta_{\rm inc})/\partial\theta_{\rm inc}>0$ for
  at least one angle $\theta_{\rm inc}$ between $0$ and $\theta_{\rm
  inc}^{\rm max}$.  Region \textit{C} contains Region \textit{B}, and
  is illustrated in Fig.\ {\ref{fig:regions}} as the area under the
  heavy solid line and to the left of the dashed line (blue in the
  color version).

\end{itemize}

Orbits in any of these three regions give possible generalizations of
the nearly horizon-skimming circular orbits presented in
Ref.~\cite{skimming}.  Notice, as illustrated in Fig.\
{\ref{fig:regions}}, that the size of these regions depends rather
strongly on the spin of the black hole.  All three regions disappear
altogether for $a < 0.9524M$ (in agreement with~\cite{skimming});
their sizes grow with $a$, reaching maximal extent for $a=M$.  These
regions never extend beyond $p \simeq 2M$.

As we shall see, the difference between these three regions is not
terribly important for assessing whether there is a strong signature
of the nearly horizon-skimming regime on the inspiral dynamics.  As
such, it is perhaps most useful to use Region \textit{C} as our
definition, since it is the most inclusive.

\section{\label{sec:iota_evolution_circular}Evolution of ${\boldsymbol
    \theta_{\rm inc}}$: circular orbits}

To ascertain whether nearly horizon-skimming orbits can affect an EMRI
in such a way as to leave a clear imprint in the gravitational-wave
signal, we have studied the time evolution of the inclination angle
$\theta_{\rm inc}$.  
To this purpose we have used the so-called adiabatic 
approximation~\cite{Mino_adiabatic}, in which the infalling body moves
along a geodesic with slowly changing parameters. The evolution of
the orbital parameters is computed using the time-averaged fluxes
$dE/dt$, $d{L}_z/dt$ and $dQ/dt$ due to gravitational-wave emission
(``radiation reaction''). As discussed in Sec.\
\ref{sec:constants_of_motion}, $E$, $L_z$ and $Q$ can be expressed in
terms of $p$, $e$, and $\theta_{\rm inc}$.  Given rates of change of
$E$, $L_z$ and $Q$, it is then straightforward
{\cite{kludge_hughes}} to calculate $dp/dt$, $de/dt$, and
$d\theta_{\rm inc}/dt$ (or $d\iota/dt$).

We should note that although perfectly well-behaved for all bound stable geodesics, 
the adiabatic approximation breaks down in a small region of the orbital parameters space 
very close to the separatrix, where the transition from an inspiral
to a plunging orbit takes place~\cite{Thorne_Ori}. However, since this
region is expected to be very small\footnote{Its width in $p/M$ is expected 
to be of the order of $\Delta p/M\sim (\mu/M)^{2/5}$, where $\mu$ is the mass of the infalling body~\cite{Thorne_Ori}.} 
and its impact on LISA waveforms rather hard 
to detect~\cite{Thorne_Ori}, we expect our results to be at least qualitatively
correct also in this region of the space of parameters.  

Accurate calculation of $dE/dt$ and $d{L}_z/dt$ in the adiabatic approximation involves solving the
Teukolsky and Sasaki-Nakamura equations {\cite{teukolsky,sasaknak}}.
For generic orbits this has been done for the first time in Ref.\
\cite{hughes_drasco}.  The calculation of $dQ/dt$ for generic orbits
is more involved.  A formula for $dQ/dt$ has been recently
derived~\cite{sago}, but has not yet been implemented (at least in a
code to which we have access).

On the other hand, it is well-known that a circular orbit will remain
circular under radiation reaction
{\cite{ryan_circ,kennefick_ori,mino_circ}}.  This constraint means
that Teukolsky-based fluxes for $E$ and $L_z$ are sufficient to
compute $dQ/dt$.  Considering this limit, the rate of change $dQ/dt$
can be expressed in terms of $dE/dt$ and $dL/dt$ as
\begin{equation}
\left(\frac{dQ}{dt}\right)_{\rm circ}\!\!\!\!\!\!\!=
-\frac{N_1(p,\iota)}{N_5(p,\iota)}\left(\frac{dE}{dt}\right)_{\rm circ}\!\!\!
-\frac{N_4(p,\iota)}{N_5(p,\iota)}\left(\frac{dL_z}{dt}\right)_{\rm
  circ}
\label{circ_relation}
\end{equation}
where 
\begin{align}
&N_{1}(p,\iota) \equiv
  E(p,\iota)\,p^{4}+a^{2}\,E(p,\iota)\,p^{2}\nonumber\\&
  \qquad\qquad\qquad\qquad-2\,a\,M\,(L_{z}(p,\iota)-a\,E(p,\iota))\,p\;,\label{N1}\\
&N_4(p,\iota) \equiv (2\,M\,p-p^2)\,L_z(p,\iota) - 2\,M\,a\,E(p,\iota)\,p\;,\label{N4}\\ 
& N_5 (p,\iota) \equiv (2\,M\,p-p^2-a^2)/2 \;.\label{N5}
\end{align}
(These quantities are for a circular orbit of radius $p$.)  Using
this, it is simple to compute $d \theta_{\rm inc}/dt$ (or
$d\iota/dt$).

This procedure was followed in Ref.~\cite{skimming}, using the code
presented in Ref.~\cite{first_code}, to determine the evolution of
$\iota$; this analysis indicated that $d\iota/dt<0$ for circular
nearly horizon-skimming orbits.  As a first step to our more general
analysis, we have repeated this calculation but using the improved
Sasaki-Nakamura-Teukolsky code presented in Ref.~\cite{hughes_drasco};
we focused on the case $a=0.998M$.

Rather to our surprise, we discovered that the fluxes $dE/dt$ and
$d{L}_z/dt$ computed with this more accurate code indicate that
$d\iota/dt>0$ (and $d\theta_{\rm inc}/dt>0$) for {\it all} circular
nearly horizon-skimming orbits --- in stark contrast with what was
found in Ref.~\cite{skimming}.  As mentioned in the introduction, the
rate of change of inclination angle appears as the difference of two
quantities.  These quantities nearly cancel (and indeed cancel exactly
in the limit $a = 0$); as such, small relative errors in their values
can lead to large relative error in the inferred inclination
evolution.  Values for $dE/dt$, $d{L}_z/dt$, $d\iota/dt$, and
$d\theta_{\rm inc}/dt$ computed using the present code are shown in
Table~\ref{teuk_circ} in the columns with the header ``Teukolsky''.

\section{\label{sec:iota_evolution_eccentric}Evolution of
  ${\boldsymbol \theta_{\rm inc}}$: non-circular orbits}

The corrected behavior of circular nearly horizon-skimming orbits has
naturally led us to investigate the evolution of non-circular nearly
horizon-skimming orbits.  Since our code cannot be used to compute
$dQ/dt$, we have resorted to a ``kludge'' approach, based on those
described in Refs.\ \cite{kludge_hughes,GG_kludge_fluxes}.  In
particular, we mostly follow the procedure developed by Gair \&
Glampedakis {\cite{GG_kludge_fluxes}}, though (as described below)
importantly modified.

The basic idea of the ``kludge'' procedure is to use the functional
form of 2PN fluxes $E$, $L_z$ and $Q$, but to correct the circular
part of these fluxes using fits to circular Teukolsky data.  As
developed in Ref.\ {\cite{GG_kludge_fluxes}}, the fluxes are written
\begin{multline}
\left(\frac{dE}{dt}\right)_{\rm GG}\hskip-0.25cm = (1-e^2)^{3/2}
\Big[(1-e^2)^{-3/2}\left(\frac{dE}{dt}\right)_{\rm
    2PN}\hskip-0.5cm(p,e,\iota)\\ - 
  \left(\frac{dE}{dt}\right)_{\rm 2PN}\hskip-0.5cm(p,0,\iota) +
  \left(\frac{dE}{dt}\right)_{\rm fit\, circ}\hskip-0.75cm(p,\iota) \Big]\;,
\label{EdotGG}
\end{multline}

\begin{multline}
\left(\frac{dL_z}{dt}\right)_{\rm GG}\hskip-0.25cm =
(1-e^2)^{3/2}
\Big[(1-e^2)^{-3/2}\left(\frac{dL_z}{dt}\right)_{\rm
    2PN}\hskip-0.5cm(p,e,\iota)\\ 
  -\left(\frac{dL_z}{dt}\right)_{\rm 2PN}\hskip-0.5cm(p,0,\iota) +
  \left(\frac{dL_z}{dt}\right)_{\rm fit\, circ}\hskip-0.75cm(p,\iota) \Big]\;,
\label{LdotGG}
\end{multline}

\begin{multline}
\left(\frac{dQ}{dt}\right)_{\rm GG}\hskip-0.25cm= (1-e^2)^{3/2}
\sqrt{Q(p,e,\iota)}\;\times \\
\Bigg[(1-e^2)^{-3/2}\left(\frac{dQ/dt}{\sqrt{Q}} \right)_{\rm
    2PN}\hskip-0.5cm(p,e,\iota) - 
\left(\frac{dQ/dt}{\sqrt{Q}} \right)_{\rm 2PN}\hskip-0.5cm(p,0,\iota) \\ +\left(\frac{dQ/dt}{\sqrt{Q}}
\right)_{\rm fit\, circ}\hskip-0.75cm(p,\iota) \Bigg]\;.
\label{QdotGG}
\end{multline}
The post-Newtonian fluxes $({dE}/{dt})_{\rm 2PN}$, $({dL_z}/{dt})_{\rm
2PN}$ and $({dQ}/{dt})_{\rm 2PN}$ are given in the Appendix
[particularly Eqs.\ (\ref{Edot2PN}), (\ref{Ldot2PN}), and
(\ref{Qdot2PN})].

Since for circular orbits the fluxes $dE/dt$, $dL_z/dt$ and $dQ/dt$
are related through Eq.\ (\ref{circ_relation}), only two fits to
circular Teukolsky data are needed.  One possible choice is to fit
$dL_z/dt$ and $d\iota/dt$, and then use the circularity constraint to
obtain\footnote{This choice might seem more involved than fitting
directly $dL_z/dt$ and $dQ/dt$, but, as noted by Gair \& Glampedakis,
ensures more sensible results for the evolution of the inclination
angle.  This generates more physically realistic inspirals
{\cite{GG_kludge_fluxes}}.}  {\cite{GG_kludge_fluxes}}
\begin{multline}
\left(\frac{dQ/dt}{\sqrt{Q}}
\right)_{\rm fit\, circ}\hskip-0.75cm(p,\iota)=\\2\,\tan\iota\left[\left(\frac{dL_z}{dt}\right)_{\rm fit\,circ}\hskip-0.75cm 
+\frac{\sqrt{Q(p,0,\iota)}}{\sin^2\iota}\left(\frac{d\iota}{dt}\right)_{\rm fit\,circ}\right]\;,
\end{multline}
\begin{multline}
  \left(\frac{dE}{dt}\right)_{\rm fit\, circ}\hskip-0.75cm
  (p,\iota)=-\frac{N_{4}(p,\iota)}{N_{1}(p,\iota)}\,\left(\frac{dL_z}{dt}\right)_{\rm
    fit\, circ}\hskip-0.75cm (p,\iota)\\-
  \frac{N_{5}(p,\iota)}{N_{1}(p,\iota)}\sqrt{Q(p,0,\iota)}\,\left(\frac{dQ/dt}{\sqrt{Q}}
  \right)_{\rm fit\, circ}\hskip-0.75cm (p,\iota)\;.
\end{multline}

As stressed in Ref.~\cite{GG_kludge_fluxes}, one does not expect these
fluxes to work well in the strong field, both because the
post-Newtonian approximation breaks down close to the black hole, and
because the circular Teukolsky data used for the fits in Ref.\
{\cite{GG_kludge_fluxes}} was computed for $3M \le p \le 30M$.  As a
first attempt to improve their behavior in the nearly horizon-skimming
region, we have made fits using circular Teukolsky data for orbits
with $M < p \leq 2M$.  In particular, for a black hole with $a =
0.998M$, we computed the circular Teukolsky-based fluxes $dL_z/dt$ and
$d\iota/dt$ listed in Table~\ref{teuk_circ} (columns 8 and 10).  These
results were fit (with error $\lesssim 0.2$\%); see Eqs.\
(\ref{LdotFitCirc}) and (\ref{IdotFitCirc}) in the Appendix.

Despite using strong-field Teukolsky fluxes for our fit, we found
fairly poor behavior of these rates of change, particularly as a
function of eccentricity.  To compensate for this, we introduced a
kludge-type fit to correct the equatorial part of the flux, in
addition to the circular part.  We fit, as a function of $p$ and $e$,
Teukolsky-based fluxes for $dE/dt$ and $dL_z/dt$ for orbits in the
equatorial plane, and then introduce the following kludge fluxes for
$E$ and $L_z$:
\begin{align}
  \frac{dE}{dt}(p,e,\iota)& =\left(\frac{dE}{dt}\right)_{\rm
    GG}\hskip-0.25cm(p,e,\iota) \nonumber\\ & -\left(\frac{dE}{dt}\right)_{\rm
    GG}\hskip-0.25cm(p,e,0)+\left(\frac{dE}{dt}\right)_{\rm
    fit\,eq}\hskip-0.5cm(p,e)\label{my_kludge_E}
  \\
  \frac{dL_z}{dt}(p,e,\iota)& =\left(\frac{dL_z}{dt}\right)_{\rm
    GG}\hskip-0.25cm(p,e,\iota) \nonumber\\& -\left(\frac{dL_z}{dt}\right)_{\rm
    GG}\hskip-0.25cm(p,e,0)+\left(\frac{dL_z}{dt}\right)_{\rm
    fit\,eq}\hskip-0.5cm(p,e)\;.\label{my_kludge_L}
\end{align}
[Note that Eq.\ (\ref{QdotGG}) for $dQ/dt$ is not modified by this
procedure since $dQ/dt=0$ for equatorial orbits.]  Using equatorial
non-circular Teukolsky data provided by Drasco {\cite{hughes_drasco,data}}
for $a = 0.998$ and $M < p \leq 2M$ (the $\iota = 0$ ``Teukolsky''
data in Tables \ref{teuk_ecc1}, \ref{teuk_ecc2} and \ref{teuk_ecc3}),
we found fits (with error $\lesssim 1.5$\%); see Eqs.\
(\ref{EdotFitEq}) and (\ref{LdotFitEq}).  Note that the 
fits for equatorial fluxes are significantly less accurate than the fits
for circular fluxes.  This appears to be due to the fact that, close
to the black hole, {\it many} harmonics are needed in order for the
Teukolsky-based fluxes to converge, especially for eccentric orbits
(cf.\ Figs.\ 2 and 3 of Ref.\ {\cite{hughes_drasco}}, noting the
number of radial harmonics that have significant contribution to the
flux).  Truncation of these sums is likely a source of some error in
the fluxes themselves, making it difficult to make a fit of as high
quality as we could in the circular case.

These fits were then finally used in Eqs.\ (\ref{my_kludge_E}) and
(\ref{my_kludge_L}) to calculate the kludge fluxes $dE/dt$ and
$dL_z/dt$ for generic orbits.  This kludge reproduces to high accuracy
our fits to the Teukolsky-based fluxes for circular orbits ($e = 0$)
or equatorial orbits ($\iota = 0$).  Some residual error remains
because the $\iota = 0$ limit of the circular fits do not precisely
equal the $e = 0$ limit of the equatorial fits.

Table {\ref{teuk_circ}} compares our kludge to Teukolsky-based fluxes
for circular orbits; the two methods agree to several digits.  Tables
\ref{teuk_ecc1}, \ref{teuk_ecc2} and \ref{teuk_ecc3} compare our
kludge to the generic Teukolsky-based fluxes for $dE/dt$ and
$dL_z/dt$ provided by Drasco {\cite{hughes_drasco,data}}.  
In all cases, the kludge fluxes $dE/dt$ and
$dL_z/dt$ have the correct qualitative behavior, being negative for
all the orbital parameters under consideration ($a=0.998 M$, $1 < p/M
\leq 2$, $0 \leq e \leq 0.5$ and $0^\circ \leq \iota \leq 41^\circ$).
The relative difference between the kludge and Teukolsky fluxes is
always less than $25$\% for $e=0$ and $e=0.1$ (even for orbits very
close to separatrix).  The accuracy remains good at larger
eccentricity, though it degrades somewhat as orbits come close to the
separatrix.

Tables \ref{teuk_circ}, \ref{teuk_ecc1}, \ref{teuk_ecc2} and
\ref{teuk_ecc3} also present the kludge values of the fluxes
$d\iota/dt$ and $d\theta_{\rm inc}/dt$ as computed using Eqs.\
(\ref{my_kludge_E}) and (\ref{my_kludge_L}) for $dE/dt$ and $dL_z/dt$,
plus Eq.\ (\ref{QdotGG}) for $dQ/dt$.  Though certainly not the last
word on inclination evolution (pending rigorous computation of
$dQ/dt$), these rates of change probably represent a better
approximation than results published to date in the literature.
(Indeed, prior work has often used the crude approximation $d\iota/dt
=0$ \cite{hughes_drasco} to estimate $dQ/dt$ given $dE/dt$ and
$dL_z/dt$.)

Most significantly, we find that $(d\iota/dt)_{\rm kludge}>0$ and
$(d\theta_{\rm inc}/dt)_{\rm kludge}>0$ for all of the orbital
parameters we consider.  In other words, we find that $d\iota/dt$ and
$d\theta_{\rm inc}/dt$ do not ever change sign. 

Finally, in Table~\ref{delta_inc} we compute the changes in
$\theta_{\rm inc}$ and $\iota$ for the inspiral with mass ratio
$\mu/M=10^{-6}$.  In all cases, we start at $p/M=1.9$.  The small body
then inspirals through the nearly horizon-skimming region until it
reaches the separatrix; at this point, the small body will fall into
the large black hole on a dynamical timescale $\sim M$, so we
terminate the calculation.  The evolution of circular orbits is
computed using our fits to the circular-Teukolsky fluxes of $E$ and
$L_z$; for eccentric orbits we use the kludge fluxes {\eqref{QdotGG}},
{\eqref{my_kludge_E}} and {\eqref{my_kludge_L}}.  As this exercise
demonstrates, the change in inclination during inspiral is never
larger than a few degrees.  Not only is there no unique sign change in
the nearly horizon-skimming region, but the magnitude of the
inclination change remains puny.  This leaves little room for the
possibility that this class of orbits may have a clear observational
imprint on the EMRI-waveforms to be detected by LISA.

\section{\label{sec:conclusions}Conclusions}

We have performed a detailed analysis of the orbital motion near the
horizon of near-extremal Kerr black holes.  We have demonstrated the
existence of a class of orbits, which we have named ``non-circular
nearly horizon-skimming orbits'', for which the angular momentum $L_z$
increases with the orbit's inclination, while keeping latus rectum and
eccentricity fixed.  This behavior, in stark contrast to that of
Newtonian orbits, generalizes earlier results for circular orbits
\cite{skimming}.

Furthermore, to assess whether this class of orbits can produce a
unique imprint on EMRI waveforms (an important source for future LISA
observations), we have studied, in the adiabatic approximation, the radiative evolution of inclination
angle for a small body orbiting in the nearly horizon-skimming region.
For circular orbits, we have re-examined the analysis of Ref.\
\cite{skimming} using an improved code for computing Teukolsky-based
fluxes of the energy and angular momentum.  Significantly correcting
Ref.\ {\cite{skimming}}'s results, we found {\it no} decrease in the
orbit's inclination angle.  Inclination always increases during
inspiral.

We next carried out such an analysis for eccentric nearly
horizon-skimming orbits. In this case, we used ``kludge'' fluxes to
evolve the constants of motion $E$, $L_z$ and $Q$
{\cite{GG_kludge_fluxes}}.  We find that these fluxes are fairly
accurate when compared with the available Teukolsky-based fluxes,
indicating that they should provide at least qualitatively correct information
regarding inclination evolution.  As for circular orbits, we find that
the orbit's inclination never decreases.  For both circular and
non-circular configurations, we find that the magnitude of the
inclination change is quite paltry --- only a few degrees at most.

Quite generically, therefore, we found that the inclination angle of
both circular and eccentric nearly horizon-skimming orbits never
decreases during the inspiral.  Revising the results obtained in Ref.\
\cite{skimming}, we thus conclude that such orbits are not likely to
yield a peculiar, unique imprint on the EMRI-waveforms detectable by
LISA.

\begin{acknowledgments}
  It is a pleasure to thank Kostas Glampedakis for enlightening
  comments and advice, and Steve Drasco for useful discussions and for
  also providing the non-circular Teukolsky data that we used in this
  paper. The supercomputers used in this investigation were provided
  by funding from the JPL Office of the Chief Information Officer.
  This work was supported in part by the DFG grant SFB TR/7, by NASA
  Grant NNG05G105G, and by NSF Grant PHY-0449884.  SAH gratefully
  acknowledges support from the MIT Class of 1956 Career Development
  Fund.
\end{acknowledgments}

\appendix* 
\section{}

In this Appendix we report the expressions for the post-Newtonian
fluxes and the fits to the Teukolsky data necessary to compute the
kludge fluxes introduced in Sec.\ \ref{sec:iota_evolution_eccentric}.
In particular the $2$PN fluxes are given by~\cite{GG_kludge_fluxes}
\begin{widetext}
\begin{eqnarray}
&&\left(\frac{dE}{dt}\right)_{\rm 2PN} = 
-\frac{32}{5} \frac{\mu^2}{M^2} \left(\frac{M}{p}\right)^5 
(1-e^2)^{3/2}\left [ g_1(e) -\tilde{a}\left(\frac{M}{p}\right)^{3/2} g_2(e)
\cos\iota -\left(\frac{M}{p}\right) g_3(e) +
\pi\left(\frac{M}{p}\right)^{3/2} g_4(e)  \right.\nonumber \\
&& \hskip 2.0cm
\left. -\left(\frac{M}{p}\right)^2 g_5(e) 
+  \tilde{a}^2\left(\frac{M}{p}\right)^2 g_6(e)
-\frac{527}{96} \tilde{a}^2 \left( \frac{M}{p}\right)^2 \sin^2\iota
\right] \;, \label{Edot2PN} \\ \nonumber \\
&&\left(\frac{dL_z}{dt}\right)_{\rm 2PN} =
-\frac{32}{5} \frac{\mu^2}{M} \left(\frac{M}{p}\right)^{7/2}
(1-e^2)^{3/2}
\left [ g_9(e)\cos\iota  
+ \tilde{a}\left(\frac{M}{p}\right)^{3/2} (g_{10}^{a}(e) 
-\cos^2\iota g_{10}^{b}(e) ) \right.
-\left(\frac{M}{p}\right) g_{11}(e) \cos\iota \nonumber\\
&& \hskip 2.2cm + \pi\left(\frac{M}{p}\right)^{3/2} g_{12}(e)\cos\iota 
-\left(\frac{M}{p}\right)^2 g_{13}(e) \cos\iota  
+ \tilde{a}^2\left(\frac{M}{p}\right)^2 \,\cos\iota\left. 
\left(g_{14}(e) - \frac{45}{8}\,\sin^2\iota\right) \right
] \;,\label{Ldot2PN} \\ \nonumber \\
&&\left(\frac{dQ}{dt}\right)_{\rm 2PN} = 
-\frac{64}{5} \frac{\mu^2}{M} \left(\frac{M}{p}\right)^{7/2} 
\,\sqrt{Q}\,\sin\iota\, (1-e^2)^{3/2}
\left[ g_9(e) - \tilde{a}\left(\frac{M}{p}\right)^{3/2} 
\cos\iota g_{10}^{b}(e) -\left(\frac{M}{p}\right)
g_{11}(e)\right. \nonumber \\ 
&&\hskip 2.0cm + \pi\left(\frac{M}{p}\right)^{3/2} g_{12}(e) 
-\left(\frac{M}{p}\right)^2 g_{13}(e)
 + \tilde{a}^2\left(\frac{M}{p}\right)^2 
\left.\times\left(g_{14}(e) -
\frac{45}{8}\,\sin^2\iota\right)\right]\;, 
\label{Qdot2PN}
\end{eqnarray}
where $\mu$ is the mass of the infalling body and where the various $e$-dependent coefficients are
\begin{align*}
&g_1(e) \equiv 1 + \frac{73}{24} e^2  + \frac{37}{96}e^4\;,
\quad
g_2(e) \equiv  \frac{73}{12} + \frac{823}{24} e^2 + \frac{949}{32}e^4
+ \frac{491}{192}e^6 \;,
\quad
g_3(e) \equiv \frac{1247}{336} + \frac{9181}{672} e^2\;,\\
&g_4(e)  \equiv 4 + \frac{1375}{48} e^2  \;,
\quad
g_5(e) \equiv \frac{44711}{9072} + \frac{172157}{2592} e^2 \;,
\quad
g_6(e) \equiv \frac{33}{16} + \frac{359}{32} e^2  \;,
\quad
g_9(e) \equiv 1 + \frac{7}{8} e^2   \;,\\
&g_{10}^{a}(e) \equiv \frac{61}{24} + \frac{63}{8}e^2   + \frac{95}{64}e^4 \;,
\quad
g_{10}^{b}(e) \equiv \frac{61}{8} + \frac{91}{4}e^2  + \frac{461}{64}e^4 \;,
\quad
g_{11}(e) \equiv \frac{1247}{336} + \frac{425}{336} e^2 \;, \\
&g_{12}(e) \equiv 4 + \frac{97}{8} e^2 \;, 
\quad
g_{13}(e) \equiv \frac{44711}{9072} + \frac{302893}{6048} e^2 \;,
\quad
g_{14}(e) \equiv \frac{33}{16} + \frac{95}{16} e^2 \;,
\end{align*}
The fits to the circular-Teukolsky data of Table~\ref{teuk_circ} are
instead given by
\begin{align}
&\left(\frac{dL_z}{dt}\right)_{\rm fit\,
    circ}
(p,\iota\,)=-\frac{32}{5} \frac{\mu^2}{M}
  \left(\frac{M}{p}\right)^{7/2} \Bigg\{ \cos\iota +
  \left(\frac{M}{p}\right)^{3/2} \left(\frac{61}{24}
  -\frac{61}{8}\,\cos^2\iota + 4\pi \cos\iota\right)
  -\frac{1247}{336}\left(\frac{M}{p}\right) \cos\iota \nonumber\\& 
  +\left(\frac{M}{p}\right)^2 \,\cos\iota \left(-\frac{1625}{567} -
  \frac{45}{8}\,\sin^2\iota\right) +
  \left(\frac{M}{p}\right)^{\frac{5}{2}}
  \Bigg[\widetilde{d}_1(p/M)+\widetilde{d}_2(p/M)\cos\iota\,+
    \widetilde{d}_3(p/M)\cos^2\iota\nonumber\\&
    +\widetilde{d}_4(p/M)\cos^3\iota\,+
    \widetilde{d}_5(p/M)\cos^4\iota+\widetilde{d}_6(p/M)\cos^5\iota+
    \cos\iota\,\left(\frac{M}{p}\right)^{3/2}\left(A+B\cos^2\iota\,
    \right)\Bigg]\Bigg\}\;,\label{LdotFitCirc} \\
\end{align}

\begin{align}
  &\left(\frac{d\iota}{dt}\right)_{\rm fit\,
    circ}
(p,\iota\,)=\frac{32}{5}\frac{\mu^2}{M}\frac{\sin^2\iota}
  {\sqrt{Q(p,0,\iota)}}\left(\frac{M}{p}\right)^{5}
  \Bigg\{\frac{61}{24}+\left(\frac{M}{p}\right)\,\widetilde{d}_1(p/M)+
  \cos\iota\,\left(\frac{M}{p}\right)^{1/2}\times\nonumber\\&
  \left[a_d^7+b_d^7\,\left(\frac{M}{p}\right)+c_d^7\,
    \left(\frac{M}{p}\right)^{3/2}\right]
  +\cos^2\iota\,\left(\frac{M}{p}\right)\,\widetilde{d}_8(p/M)+
  \,\cos\iota\,\left(\frac{M}{p}\right)^{5/2}
  \left[\widetilde{h}_1(p/M)+\cos^2\iota\,
    \widetilde{h}_2(p/M)\right]\Bigg\}\;,\label{IdotFitCirc}
\end{align}
where
\begin{gather}
  \widetilde{d}_i(x) \equiv a_d^i+b_d^i\,x^{-1/2}+c_d^i\,x^{-1}\;, \qquad
  i=1,\ldots,8,
  \qquad \qquad \widetilde{h}_i(x)\equiv a_h^i+b_h^i\,x^{-1/2}\;,
  \quad i=1,2
\end{gather}
and the numerical coefficients are given by
\begin{equation}
a_h^1=-278.9387 \;,\quad b_h^1=84.1414\;,\quad
a_h^2=8.6679 \;,\quad  b_h^2=-9.2401\;,\quad
A= -18.3362 \;,\quad B= 24.9034 \;,
 \end{equation}
and by the following table
\begin{table}[h]
\begin{tabular}{|c|c|c|c|c|c|c|c|c|}
\hline
\T \B$i$ & \B1 &\B2 &\B3 &\B4 &\B5 &\B6 &\B7 &\B8\\
\hline
\T \B$a_d^i$ & 15.8363 & 445.4418 & $-$2027.7797 & 3089.1709 & $-$2045.2248 & 498.6411 & $-$8.7220 & 50.8345 \\
\T \B$b_d^i$ & $-$55.6777 & $-$1333.2461 & 5940.4831 & $-$9103.4472 & 6113.1165 & $-$1515.8506 & $-$50.8950 & $-$131.6422 \\
\T \B$c_d^i$ & 38.6405  & 1049.5637 & $-$4513.0879 & 6926.3191 &-4714.9633 & 1183.5875 & 251.4025& 83.0834\\
\hline
\end{tabular}
\label{table:a_d_i_coefficients}
\end{table}

Note that the functional form of these fits was obtained from
Eqs.\ (57) and (58) of Ref.~\cite{GG_kludge_fluxes} by setting
$\tilde{a}$ (\textit{i.e.}, $q$ in their notation) to $1$. Finally, we
give expressions for the fits to the equatorial Teukolsky data of
tables \ref{teuk_ecc1}, \ref{teuk_ecc2} and \ref{teuk_ecc3} (data with
$\iota=0$, columns with header ``Teukolsky''):
\begin{align}
&\left(\frac{dE}{dt}\right)_{\rm fit\, eq}\hskip-0.5cm(p,e)=
  \left(\frac{dE}{dt}\right)_{2PN}\hskip-0.5cm(p,e,0)-\frac{32}{5}\left(\frac{\mu}{M}\right)^2\left(\frac{M}{p}\right)^5
  (1-e^2)^{3/2}\Bigg[\widetilde{g}_1(e)+\widetilde{g}_2(e)\left(\frac{M}{p}\right)^{1/2}
    +\widetilde{g}_3(e)\left(\frac{M}{p}\right)\nonumber\\&
    \hskip10.cm+\widetilde{g}_4(e)\left(\frac{M}{p}\right)^{3/2}+
    \widetilde{g}_5(e)\left(\frac{M}{p}\right)^2\Bigg]\;,\label{EdotFitEq}
  \\ & \left(\frac{L_z}{dt}\right)_{\rm fit\,
    eq}\hskip-0.5cm(p,e)=\left(\frac{L_z}{dt}\right)_{2PN}\hskip-0.5cm(p,e,0)
  -\frac{32}{5}\frac{\mu^2}{M}\left(\frac{M}{p}\right)^{7/2}
  (1-e^2)^{3/2}\Bigg[\widetilde{f}_1(e)+\widetilde{f}_2(e)\left(\frac{M}{p}\right)^{1/2}+\widetilde{f}_3(e)\left(\frac{M}{p}\right)
    \nonumber\\&
    \hskip9.5cm+\widetilde{f}_4(e)\left(\frac{M}{p}\right)^{3/2}+
    \widetilde{f}_5(e)\left(\frac{M}{p}\right)^2\Bigg]\;,\label{LdotFitEq}
\end{align}
\begin{equation}
\widetilde{g}_i(e) \equiv a_g^i+b_g^i\,e^2+c_g^i\,e^4+d_g^i\,e^6\;, \qquad \qquad 
\widetilde{f}_i(e) \equiv a_f^i+b_f^i\,e^2+c_f^i\,e^4+d_f^i\,e^6\;, \quad \quad  i=1,\ldots,5
\end{equation}
where the numerical coefficients are given by the following table
\begin{table}[h]
\begin{tabular}{|c|c|c|c|c|c|c|c|c|}
\hline
\T \B$i$ & $a_g^i$ & $b_g^i$ & $c_g^i$ & $d_g^i$ & $a_f^i$ & $b_f^i$ & $c_f^i$& $d_f^i$ \\
\hline
\T \B 1  & 6.4590 & $-$2038.7301 & 6639.9843 & 227709.2187 & 5.4577&  $-$3116.4034& 4711.7065 & 214332.2907\\
\T \B 2 &-31.2215 & 10390.6778& $-$27505.7295& $-$1224376.5294 &    $-$26.6519 & 15958.6191 & $-$16390.4868 & $-$1147201.4687\\
\T \B 3 &57.1208& $-$19800.4891& 39527.8397& 2463977.3622 &    50.4374 &-30579.3129& 15749.9411 & 2296989.5466\\
\T \B 4 &-49.7051& 16684.4629& $-$21714.7941& $-$2199231.9494 &     $-$46.7816 & 25968.8743 & 656.3460 & $-$2038650.9838\\
\T \B 5 &16.4697& $-$5234.2077& 2936.2391&  734454.5696 &    15.6660 & $-$8226.3892 & $-$4903.9260 & 676553.2755\\
\hline
\end{tabular}
\end{table}

\begin{table}[h]
\begin{tabular}{|c|c|c|c|c|c|c|c|c|c|c|c|}
\hline
 \T\B$\frac{p}{M}$ & $e$ & $\theta_{\rm inc}$& $\iota$& $\frac{dE}{dt}\times\frac{M^2}{\mu^2}$ & $\frac{dE}{dt}\times\frac{M^2}{\mu^2}$ & $\frac{dL_z}{dt}\times\frac{M}{\mu^2}$ & $\frac{dL_z}{dt}\times\frac{M}{\mu^2}$ & $\frac{d\iota}{dt}\times\frac{M}{\mu^2}$& $\frac{d\iota}{dt}\times\frac{M}{\mu^2}$& $\frac{d\theta_{\rm inc}}{dt}\times\frac{M}{\mu^2}$& $\frac{d\theta_{\rm inc}}{dt}\times\frac{M}{\mu^2}$\\
&&(deg.)&(deg.)& (kludge) & (Teukolsky) & (kludge) & (Teukolsky)& (kludge) & (Teukolsky)& (kludge) & (Teukolsky)\\
\hline
\T 

  1.3&  0&  0&  0& $-$9.108$\times$10$^{-2}$& $-$9.109$\times$10$^{-2}$& $-$2.258$\times$10$^{-1}$& $-$2.259$\times$10$^{-1}$&  0&  0&  0&  0\\
  1.3&  0& 10.4870& 11.6773& $-$9.328$\times$10$^{-2}$& $-$9.332$\times$10$^{-2}$& $-$2.304$\times$10$^{-1}$& $-$2.306$\times$10$^{-1}$&  1.837$\times$10$^{-2}$&  1.839$\times$10$^{-2}$&  6.462$\times$10$^{-3}$&  6.475$\times$10$^{-3}$\\
  1.3&  0& 14.6406& 16.1303& $-$9.588$\times$10$^{-2}$& $-$9.588$\times$10$^{-2}$& $-$2.359$\times$10$^{-1}$& $-$2.360$\times$10$^{-1}$&  2.397$\times$10$^{-2}$&  2.400$\times$10$^{-2}$&  8.645$\times$10$^{-3}$&  8.667$\times$10$^{-3}$\\
  1.3&  0& 17.7000& 19.3172& $-$9.875$\times$10$^{-2}$& $-$9.876$\times$10$^{-2}$& $-$2.420$\times$10$^{-1}$& $-$2.421$\times$10$^{-1}$&  2.728$\times$10$^{-2}$&  2.731$\times$10$^{-2}$&  1.007$\times$10$^{-2}$&  1.010$\times$10$^{-2}$\\
  1.3&  0& 20.1636& 21.8210& $-$1.019$\times$10$^{-1}$& $-$1.019$\times$10$^{-1}$& $-$2.486$\times$10$^{-1}$& $-$2.488$\times$10$^{-1}$&  2.943$\times$10$^{-2}$&  2.950$\times$10$^{-2}$&  1.111$\times$10$^{-2}$&  1.117$\times$10$^{-2}$\\
  1.4&  0&  0&  0& $-$8.700$\times$10$^{-2}$& $-$8.709$\times$10$^{-2}$& $-$2.311$\times$10$^{-1}$& $-$2.312$\times$10$^{-1}$&  0&  0&  0&  0\\
  1.4&  0& 14.5992& 16.0005& $-$9.062$\times$10$^{-2}$& $-$9.070$\times$10$^{-2}$& $-$2.386$\times$10$^{-1}$& $-$2.386$\times$10$^{-1}$&  2.316$\times$10$^{-2}$&  2.319$\times$10$^{-2}$&  8.823$\times$10$^{-3}$&  8.848$\times$10$^{-3}$\\
  1.4&  0& 20.1756& 21.7815& $-$9.520$\times$10$^{-2}$& $-$9.526$\times$10$^{-2}$& $-$2.482$\times$10$^{-1}$& $-$2.482$\times$10$^{-1}$&  2.875$\times$10$^{-2}$&  2.877$\times$10$^{-2}$&  1.141$\times$10$^{-2}$&  1.143$\times$10$^{-2}$\\
  1.4&  0& 24.1503& 25.7517& $-$1.006$\times$10$^{-1}$& $-$1.007$\times$10$^{-1}$& $-$2.595$\times$10$^{-1}$& $-$2.596$\times$10$^{-1}$&  3.140$\times$10$^{-2}$&  3.141$\times$10$^{-2}$&  1.289$\times$10$^{-2}$&  1.288$\times$10$^{-2}$\\
  1.4&  0& 27.2489& 28.7604& $-$1.067$\times$10$^{-1}$& $-$1.068$\times$10$^{-1}$& $-$2.725$\times$10$^{-1}$& $-$2.725$\times$10$^{-1}$&  3.274$\times$10$^{-2}$&  3.275$\times$10$^{-2}$&  1.378$\times$10$^{-2}$&  1.377$\times$10$^{-2}$\\
  1.5&  0&  0&  0& $-$8.009$\times$10$^{-2}$& $-$7.989$\times$10$^{-2}$& $-$2.270$\times$10$^{-1}$& $-$2.265$\times$10$^{-1}$&  0&  0&  0&  0\\
  1.5&  0& 16.7836& 18.1857& $-$8.401$\times$10$^{-2}$& $-$8.383$\times$10$^{-2}$& $-$2.348$\times$10$^{-1}$& $-$2.343$\times$10$^{-1}$&  2.360$\times$10$^{-2}$&  2.351$\times$10$^{-2}$&  9.602$\times$10$^{-3}$&  9.545$\times$10$^{-3}$\\
  1.5&  0& 23.0755& 24.6167& $-$8.917$\times$10$^{-2}$& $-$8.897$\times$10$^{-2}$& $-$2.454$\times$10$^{-1}$& $-$2.449$\times$10$^{-1}$&  2.872$\times$10$^{-2}$&  2.863$\times$10$^{-2}$&  1.228$\times$10$^{-2}$&  1.222$\times$10$^{-2}$\\
  1.5&  0& 27.4892& 28.9670& $-$9.537$\times$10$^{-2}$& $-$9.516$\times$10$^{-2}$& $-$2.583$\times$10$^{-1}$& $-$2.579$\times$10$^{-1}$&  3.091$\times$10$^{-2}$&  3.082$\times$10$^{-2}$&  1.372$\times$10$^{-2}$&  1.367$\times$10$^{-2}$\\
  1.5&  0& 30.8795& 32.2231& $-$1.025$\times$10$^{-1}$& $-$1.023$\times$10$^{-1}$& $-$2.733$\times$10$^{-1}$& $-$2.728$\times$10$^{-1}$&  3.184$\times$10$^{-2}$&  3.173$\times$10$^{-2}$&  1.452$\times$10$^{-2}$&  1.443$\times$10$^{-2}$\\
  1.6&  0&  0&  0& $-$7.181$\times$10$^{-2}$& $-$7.156$\times$10$^{-2}$& $-$2.168$\times$10$^{-1}$& $-$2.162$\times$10$^{-1}$&  0&  0&  0&  0\\
  1.6&  0& 18.3669& 19.7220& $-$7.568$\times$10$^{-2}$& $-$7.545$\times$10$^{-2}$& $-$2.242$\times$10$^{-1}$& $-$2.237$\times$10$^{-1}$&  2.240$\times$10$^{-2}$&  2.229$\times$10$^{-2}$&  9.600$\times$10$^{-3}$&  9.515$\times$10$^{-3}$\\
  1.6&  0& 25.1720& 26.6245& $-$8.084$\times$10$^{-2}$& $-$8.062$\times$10$^{-2}$& $-$2.346$\times$10$^{-1}$& $-$2.341$\times$10$^{-1}$&  2.701$\times$10$^{-2}$&  2.685$\times$10$^{-2}$&  1.223$\times$10$^{-2}$&  1.210$\times$10$^{-2}$\\
  1.6&  0& 29.9014& 31.2625& $-$8.708$\times$10$^{-2}$& $-$8.687$\times$10$^{-2}$& $-$2.474$\times$10$^{-1}$& $-$2.470$\times$10$^{-1}$&  2.889$\times$10$^{-2}$&  2.872$\times$10$^{-2}$&  1.363$\times$10$^{-2}$&  1.349$\times$10$^{-2}$\\
  1.6&  0& 33.5053& 34.7164& $-$9.425$\times$10$^{-2}$& $-$9.399$\times$10$^{-2}$& $-$2.622$\times$10$^{-1}$& $-$2.616$\times$10$^{-1}$&  2.964$\times$10$^{-2}$&  2.951$\times$10$^{-2}$&  1.441$\times$10$^{-2}$&  1.432$\times$10$^{-2}$\\
  1.7&  0&  0&  0& $-$6.332$\times$10$^{-2}$& $-$6.317$\times$10$^{-2}$& $-$2.034$\times$10$^{-1}$& $-$2.031$\times$10$^{-1}$&  0&  0&  0&  0\\
  1.7&  0& 19.6910& 20.9859& $-$6.702$\times$10$^{-2}$& $-$6.687$\times$10$^{-2}$& $-$2.101$\times$10$^{-1}$& $-$2.098$\times$10$^{-1}$&  2.057$\times$10$^{-2}$&  2.052$\times$10$^{-2}$&  9.202$\times$10$^{-3}$&  9.171$\times$10$^{-3}$\\
  1.7&  0& 26.9252& 28.2884& $-$7.197$\times$10$^{-2}$& $-$7.184$\times$10$^{-2}$& $-$2.199$\times$10$^{-1}$& $-$2.196$\times$10$^{-1}$&  2.467$\times$10$^{-2}$&  2.456$\times$10$^{-2}$&  1.170$\times$10$^{-2}$&  1.162$\times$10$^{-2}$\\
  1.7&  0& 31.9218& 33.1786& $-$7.794$\times$10$^{-2}$& $-$7.782$\times$10$^{-2}$& $-$2.319$\times$10$^{-1}$& $-$2.316$\times$10$^{-1}$&  2.632$\times$10$^{-2}$&  2.620$\times$10$^{-2}$&  1.306$\times$10$^{-2}$&  1.296$\times$10$^{-2}$\\
  1.7&  0& 35.7100& 36.8118& $-$8.475$\times$10$^{-2}$& $-$8.465$\times$10$^{-2}$& $-$2.457$\times$10$^{-1}$& $-$2.455$\times$10$^{-1}$&  2.698$\times$10$^{-2}$&  2.686$\times$10$^{-2}$&  1.384$\times$10$^{-2}$&  1.373$\times$10$^{-2}$\\
  1.8&  0&  0&  0& $-$5.531$\times$10$^{-2}$& $-$5.528$\times$10$^{-2}$& $-$1.888$\times$10$^{-1}$& $-$1.887$\times$10$^{-1}$&  0&  0&  0&  0\\
  1.8&  0& 20.8804& 22.1128& $-$5.879$\times$10$^{-2}$& $-$5.874$\times$10$^{-2}$& $-$1.948$\times$10$^{-1}$& $-$1.946$\times$10$^{-1}$&  1.858$\times$10$^{-2}$&  1.858$\times$10$^{-2}$&  8.635$\times$10$^{-3}$&  8.639$\times$10$^{-3}$\\
  1.8&  0& 28.5007& 29.7791& $-$6.343$\times$10$^{-2}$& $-$6.336$\times$10$^{-2}$& $-$2.036$\times$10$^{-1}$& $-$2.035$\times$10$^{-1}$&  2.221$\times$10$^{-2}$&  2.223$\times$10$^{-2}$&  1.098$\times$10$^{-2}$&  1.101$\times$10$^{-2}$\\
  1.8&  0& 33.7400& 34.9034& $-$6.901$\times$10$^{-2}$& $-$6.894$\times$10$^{-2}$& $-$2.146$\times$10$^{-1}$& $-$2.144$\times$10$^{-1}$&  2.368$\times$10$^{-2}$&  2.371$\times$10$^{-2}$&  1.228$\times$10$^{-2}$&  1.232$\times$10$^{-2}$\\
  1.8&  0& 37.6985& 38.7065& $-$7.533$\times$10$^{-2}$& $-$7.533$\times$10$^{-2}$& $-$2.271$\times$10$^{-1}$& $-$2.271$\times$10$^{-1}$&  2.429$\times$10$^{-2}$&  2.427$\times$10$^{-2}$&  1.306$\times$10$^{-2}$&  1.303$\times$10$^{-2}$\\
  1.9&  0&  0&  0& $-$4.809$\times$10$^{-2}$& $-$4.811$\times$10$^{-2}$& $-$1.740$\times$10$^{-1}$& $-$1.740$\times$10$^{-1}$&  0&  0&  0&  0\\
  1.9&  0& 21.9900& 23.1615& $-$5.132$\times$10$^{-2}$& $-$5.134$\times$10$^{-2}$& $-$1.792$\times$10$^{-1}$& $-$1.793$\times$10$^{-1}$&  1.666$\times$10$^{-2}$&  1.664$\times$10$^{-2}$&  8.022$\times$10$^{-3}$&  8.007$\times$10$^{-3}$\\
  1.9&  0& 29.9708& 31.1702& $-$5.562$\times$10$^{-2}$& $-$5.564$\times$10$^{-2}$& $-$1.872$\times$10$^{-1}$& $-$1.872$\times$10$^{-1}$&  1.986$\times$10$^{-2}$&  1.987$\times$10$^{-2}$&  1.019$\times$10$^{-2}$&  1.020$\times$10$^{-2}$\\
  1.9&  0& 35.4385& 36.5176& $-$6.078$\times$10$^{-2}$& $-$6.077$\times$10$^{-2}$& $-$1.971$\times$10$^{-1}$& $-$1.970$\times$10$^{-1}$&  2.118$\times$10$^{-2}$&  2.122$\times$10$^{-2}$&  1.143$\times$10$^{-2}$&  1.148$\times$10$^{-2}$\\
  1.9&  0& 39.5592& 40.4847& $-$6.659$\times$10$^{-2}$& $-$6.658$\times$10$^{-2}$& $-$2.082$\times$10$^{-1}$& $-$2.082$\times$10$^{-1}$&  2.177$\times$10$^{-2}$&  2.182$\times$10$^{-2}$&  1.222$\times$10$^{-2}$&  1.228$\times$10$^{-2}$\\
  2.0&  0&  0&  0& $-$4.174$\times$10$^{-2}$& $-$4.175$\times$10$^{-2}$& $-$1.598$\times$10$^{-1}$& $-$1.598$\times$10$^{-1}$&  0&  0&  0&  0\\
  2.0&  0& 23.0471& 24.1605& $-$4.471$\times$10$^{-2}$& $-$4.472$\times$10$^{-2}$& $-$1.643$\times$10$^{-1}$& $-$1.643$\times$10$^{-1}$&  1.489$\times$10$^{-2}$&  1.489$\times$10$^{-2}$&  7.425$\times$10$^{-3}$&  7.424$\times$10$^{-3}$\\
  2.0&  0& 31.3715& 32.4978& $-$4.867$\times$10$^{-2}$& $-$4.871$\times$10$^{-2}$& $-$1.713$\times$10$^{-1}$& $-$1.714$\times$10$^{-1}$&  1.773$\times$10$^{-2}$&  1.770$\times$10$^{-2}$&  9.436$\times$10$^{-3}$&  9.411$\times$10$^{-3}$\\
  2.0&  0& 37.0583& 38.0608& $-$5.341$\times$10$^{-2}$& $-$5.345$\times$10$^{-2}$& $-$1.801$\times$10$^{-1}$& $-$1.801$\times$10$^{-1}$&  1.893$\times$10$^{-2}$&  1.889$\times$10$^{-2}$&  1.062$\times$10$^{-2}$&  1.057$\times$10$^{-2}$\\
  2.0&  0& 41.3358& 42.1876& $-$5.873$\times$10$^{-2}$& $-$5.875$\times$10$^{-2}$& $-$1.900$\times$10$^{-1}$& $-$1.900$\times$10$^{-1}$&  1.950$\times$10$^{-2}$&  1.948$\times$10$^{-2}$&  1.141$\times$10$^{-2}$&  1.138$\times$10$^{-2}$\\

 \hline
\end{tabular}
\caption{Teukolsky-based fluxes and kludge fluxes [computed using
Eqs.\ (\ref{QdotGG}), (\ref{my_kludge_E}) and (\ref{my_kludge_L})] for
circular orbits about a hole with $a=0.998 M$; $\mu$ represents the mass of the infalling body.
The Teukolsky-based fluxes have an accuracy of $10^{-6}$.
\label{teuk_circ}} 
\end{table}

\begin{table}[h]
\begin{tabular}{|c|c|c|c|c|c|c|c|c|c|}
\hline
 \T\B$\frac{p}{M}$ & $e$ & $\theta_{\rm inc}$& $\iota$& $\frac{dE}{dt}\times\frac{M^2}{\mu^2}$ & $\frac{dE}{dt}\times\frac{M^2}{\mu^2}$ & $\frac{dL_z}{dt}\times\frac{M}{\mu^2}$ & $\frac{dL_z}{dt}\times\frac{M}{\mu^2}$ & $\frac{d\iota}{dt}\times\frac{M}{\mu^2}$& $\frac{d\theta_{\rm inc}}{dt}\times\frac{M}{\mu^2}$\\
&&(deg.)&(deg.)& (kludge) & (Teukolsky) & (kludge) & (Teukolsky)& (kludge) &  (kludge) \\
\hline
\T 1.3&  0.1&  0&  0& $-$8.804$\times$10$^{-2}$& $-$8.804$\times$10$^{-2}$& $-$2.098$\times$10$^{-1}$& $-$2.098$\times$10$^{-1}$&  0&  0\\
  1.4&  0.1&  0&  0& $-$8.728$\times$10$^{-2}$& $-$8.719$\times$10$^{-2}$& $-$2.274$\times$10$^{-1}$& $-$2.275$\times$10$^{-1}$&  0&   0\\
  1.4&  0.1&  8&  8.8664& $-$9.110$\times$10$^{-2}$& $-$8.736$\times$10$^{-2}$& $-$2.355$\times$10$^{-1}$& $-$2.273$\times$10$^{-1}$&  4.066$\times$10$^{-2}$&  2.938$\times$10$^{-2}$\\
  1.4&  0.1& 16& 17.4519& $-$1.030$\times$10$^{-1}$& $-$8.958$\times$10$^{-2}$& $-$2.602$\times$10$^{-1}$& $-$2.309$\times$10$^{-1}$&  7.428$\times$10$^{-2}$&   5.475$\times$10$^{-2}$\\
  1.4&  0.1& 24& 25.5784& $-$1.243$\times$10$^{-1}$& $-$9.771$\times$10$^{-2}$& $-$3.037$\times$10$^{-1}$& $-$2.415$\times$10$^{-1}$&  9.663$\times$10$^{-2}$&  7.316$\times$10$^{-2}$\\
  1.5&  0.1&  0&  0& $-$8.069$\times$10$^{-2}$& $-$8.095$\times$10$^{-2}$& $-$2.255$\times$10$^{-1}$& $-$2.260$\times$10$^{-1}$&  0&   0\\
  1.5&  0.1&  8&  8.7910& $-$8.323$\times$10$^{-2}$& $-$8.133$\times$10$^{-2}$& $-$2.310$\times$10$^{-1}$& $-$2.264$\times$10$^{-1}$&  2.996$\times$10$^{-2}$&   2.070$\times$10$^{-2}$\\
  1.5&  0.1& 16& 17.3490& $-$9.121$\times$10$^{-2}$& $-$8.395$\times$10$^{-2}$& $-$2.483$\times$10$^{-1}$& $-$2.314$\times$10$^{-1}$&  5.512$\times$10$^{-2}$&   3.888$\times$10$^{-2}$\\
  1.5&  0.1& 24& 25.5197& $-$1.059$\times$10$^{-1}$& $-$8.980$\times$10$^{-2}$& $-$2.792$\times$10$^{-1}$& $-$2.423$\times$10$^{-1}$&  7.255$\times$10$^{-2}$&   5.264$\times$10$^{-2}$\\
  1.6&  0.1&  0&  0& $-$7.255$\times$10$^{-2}$& $-$7.281$\times$10$^{-2}$& $-$2.161$\times$10$^{-1}$& $-$2.168$\times$10$^{-1}$&  0&   0\\
  1.6&  0.1&  8&  8.7195& $-$7.430$\times$10$^{-2}$& $-$7.321$\times$10$^{-2}$& $-$2.201$\times$10$^{-1}$& $-$2.173$\times$10$^{-1}$&  2.258$\times$10$^{-2}$&   1.502$\times$10$^{-2}$\\
  1.6&  0.1& 16& 17.2437& $-$7.986$\times$10$^{-2}$& $-$7.533$\times$10$^{-2}$& $-$2.323$\times$10$^{-1}$& $-$2.212$\times$10$^{-1}$&  4.179$\times$10$^{-2}$&    2.839$\times$10$^{-2}$\\
  1.6&  0.1& 24& 25.4388& $-$9.025$\times$10$^{-2}$& $-$8.040$\times$10$^{-2}$& $-$2.547$\times$10$^{-1}$& $-$2.309$\times$10$^{-1}$&  5.554$\times$10$^{-2}$&    3.886$\times$10$^{-2}$\\
  1.6&  0.1& 32& 33.2683& $-$1.082$\times$10$^{-1}$& $-$9.435$\times$10$^{-2}$& $-$2.920$\times$10$^{-1}$& $-$2.551$\times$10$^{-1}$&  6.316$\times$10$^{-2}$&    4.559$\times$10$^{-2}$\\
  1.7&  0.1&  0&  0& $-$6.427$\times$10$^{-2}$& $-$6.440$\times$10$^{-2}$& $-$2.036$\times$10$^{-1}$& $-$2.040$\times$10$^{-1}$&  0&   0\\
  1.7&  0.1&  8&  8.6555& $-$6.552$\times$10$^{-2}$& $-$6.478$\times$10$^{-2}$& $-$2.065$\times$10$^{-1}$& $-$2.045$\times$10$^{-1}$&  1.742$\times$10$^{-2}$&    1.124$\times$10$^{-2}$\\
  1.7&  0.1& 16& 17.1454& $-$6.953$\times$10$^{-2}$& $-$6.651$\times$10$^{-2}$& $-$2.154$\times$10$^{-1}$& $-$2.075$\times$10$^{-1}$&  3.240$\times$10$^{-2}$&    2.134$\times$10$^{-2}$\\
  1.7&  0.1& 24& 25.3531& $-$7.707$\times$10$^{-2}$& $-$7.052$\times$10$^{-2}$& $-$2.317$\times$10$^{-1}$& $-$2.150$\times$10$^{-1}$&  4.342$\times$10$^{-2}$&    2.948$\times$10$^{-2}$\\
  1.7&  0.1& 32& 33.2416& $-$9.009$\times$10$^{-2}$& $-$7.959$\times$10$^{-2}$& $-$2.590$\times$10$^{-1}$& $-$2.324$\times$10$^{-1}$&  4.998$\times$10$^{-2}$&    3.512$\times$10$^{-2}$\\
  1.8&  0.1&  0&  0& $-$5.640$\times$10$^{-2}$& $-$5.640$\times$10$^{-2}$& $-$1.897$\times$10$^{-1}$& $-$1.897$\times$10$^{-1}$&  0&    0\\
  1.8&  0.1&  8&  8.5991& $-$5.732$\times$10$^{-2}$& $-$5.676$\times$10$^{-2}$& $-$1.918$\times$10$^{-1}$& $-$1.902$\times$10$^{-1}$&  1.371$\times$10$^{-2}$&    8.640$\times$10$^{-3}$\\
  1.8&  0.1& 16& 17.0562& $-$6.028$\times$10$^{-2}$& $-$5.817$\times$10$^{-2}$& $-$1.984$\times$10$^{-1}$& $-$1.925$\times$10$^{-1}$&  2.562$\times$10$^{-2}$&    1.647$\times$10$^{-2}$\\
  1.8&  0.1& 24& 25.2693& $-$6.588$\times$10$^{-2}$& $-$6.139$\times$10$^{-2}$& $-$2.105$\times$10$^{-1}$& $-$1.983$\times$10$^{-1}$&  3.456$\times$10$^{-2}$&    2.291$\times$10$^{-2}$\\
  1.8&  0.1& 32& 33.2018& $-$7.555$\times$10$^{-2}$& $-$6.849$\times$10$^{-2}$& $-$2.307$\times$10$^{-1}$& $-$2.120$\times$10$^{-1}$&  4.020$\times$10$^{-2}$&    2.765$\times$10$^{-2}$\\
  1.9&  0.1&  0&  0& $-$4.915$\times$10$^{-2}$& $-$4.911$\times$10$^{-2}$& $-$1.753$\times$10$^{-1}$& $-$1.751$\times$10$^{-1}$&  0&    0\\
  1.9&  0.1&  8&  8.5494& $-$4.985$\times$10$^{-2}$& $-$4.945$\times$10$^{-2}$& $-$1.768$\times$10$^{-1}$& $-$1.755$\times$10$^{-1}$&  1.097$\times$10$^{-2}$&    6.791$\times$10$^{-3}$\\
  1.9&  0.1& 16& 16.9760& $-$5.208$\times$10$^{-2}$& $-$5.064$\times$10$^{-2}$& $-$1.817$\times$10$^{-1}$& $-$1.774$\times$10$^{-1}$&  2.055$\times$10$^{-2}$&    1.298$\times$10$^{-2}$\\
  1.9&  0.1& 24& 25.1898& $-$5.633$\times$10$^{-2}$& $-$5.328$\times$10$^{-2}$& $-$1.908$\times$10$^{-1}$& $-$1.819$\times$10$^{-1}$&  2.788$\times$10$^{-2}$&    1.816$\times$10$^{-2}$\\
  1.9&  0.1& 32& 33.1555& $-$6.364$\times$10$^{-2}$& $-$5.870$\times$10$^{-2}$& $-$2.059$\times$10$^{-1}$& $-$1.920$\times$10$^{-1}$&  3.272$\times$10$^{-2}$&    2.214$\times$10$^{-2}$\\
  2.0&  0.1&  0&  0& $-$4.263$\times$10$^{-2}$& $-$4.264$\times$10$^{-2}$& $-$1.607$\times$10$^{-1}$& $-$1.608$\times$10$^{-1}$&  0&    0\\
  2.0&  0.1&  8&  8.5057& $-$4.316$\times$10$^{-2}$& $-$4.292$\times$10$^{-2}$& $-$1.619$\times$10$^{-1}$& $-$1.611$\times$10$^{-1}$&  8.862$\times$10$^{-3}$&    5.424$\times$10$^{-3}$\\
  2.0&  0.1& 16& 16.9042& $-$4.488$\times$10$^{-2}$& $-$4.390$\times$10$^{-2}$& $-$1.656$\times$10$^{-1}$& $-$1.625$\times$10$^{-1}$&  1.666$\times$10$^{-2}$&    1.039$\times$10$^{-2}$\\
  2.0&  0.1& 24& 25.1156& $-$4.815$\times$10$^{-2}$& $-$4.604$\times$10$^{-2}$& $-$1.724$\times$10$^{-1}$& $-$1.660$\times$10$^{-1}$&  2.271$\times$10$^{-2}$&    1.459$\times$10$^{-2}$\\
  2.0&  0.1& 32& 33.1064& $-$5.376$\times$10$^{-2}$& $-$5.031$\times$10$^{-2}$& $-$1.838$\times$10$^{-1}$& $-$1.736$\times$10$^{-1}$&  2.684$\times$10$^{-2}$&    1.793$\times$10$^{-2}$\\
  2.0&  0.1& 40& 40.8954& $-$6.339$\times$10$^{-2}$& $-$6.236$\times$10$^{-2}$& $-$2.027$\times$10$^{-1}$& $-$1.967$\times$10$^{-1}$&  2.917$\times$10$^{-2}$&    2.036$\times$10$^{-2}$\\
\hline
\end{tabular}
\caption{As in Table~\ref{teuk_circ} but for non-circular orbits;
the Teukolsky-based fluxes for $E$ and $L_z$ have an accuracy of $10^{-3}$.
Note that our code, as all the Teukolsky-based code that we are aware of, 
presently does not have the capability to compute inclination angle evolution for generic
orbits.\label{teuk_ecc1}} 
\end{table}

\begin{table}[h]
\begin{tabular}{|c|c|c|c|c|c|c|c|c|c|}
\hline
 \T\B$\frac{p}{M}$ & $e$ & $\theta_{\rm inc}$& $\iota$& $\frac{dE}{dt}\times\frac{M^2}{\mu^2}$ & $\frac{dE}{dt}\times\frac{M^2}{\mu^2}$ & $\frac{dL_z}{dt}\times\frac{M}{\mu^2}$ & $\frac{dL_z}{dt}\times\frac{M}{\mu^2}$ & $\frac{d\iota}{dt}\times\frac{M}{\mu^2}$&  $\frac{d\theta_{\rm inc}}{dt}\times\frac{M}{\mu^2}$\\
&&(deg.)&(deg.)& (kludge) & (Teukolsky ) & (kludge) & (Teukolsky)& (kludge) & (kludge) \\
\hline
\T 1.4&  0.2&  0&  0& $-$8.636$\times$10$^{-2}$& $-$8.642$\times$10$^{-2}$& $-$2.119$\times$10$^{-1}$& $-$2.121$\times$10$^{-1}$&  0&    0\\
  1.4&  0.2&  8&  8.8215& $-$9.853$\times$10$^{-2}$& $-$8.240$\times$10$^{-2}$& $-$2.374$\times$10$^{-1}$& $-$2.015$\times$10$^{-1}$&  1.148$\times$10$^{-1}$&   9.714$\times$10$^{-2}$\\
  1.5&  0.2&  0&  0& $-$8.362$\times$10$^{-2}$& $-$8.349$\times$10$^{-2}$& $-$2.236$\times$10$^{-1}$& $-$2.230$\times$10$^{-1}$&  0&    0\\
  1.5&  0.2&  8&  8.7595& $-$9.141$\times$10$^{-2}$& $-$8.276$\times$10$^{-2}$& $-$2.410$\times$10$^{-1}$& $-$2.206$\times$10$^{-1}$&  7.893$\times$10$^{-2}$&    6.549$\times$10$^{-2}$\\
  1.5&  0.2& 16& 17.2957& $-$1.145$\times$10$^{-1}$& $-$8.394$\times$10$^{-2}$& $-$2.915$\times$10$^{-1}$& $-$2.215$\times$10$^{-1}$&  1.466$\times$10$^{-1}$&    1.230$\times$10$^{-1}$\\
  1.5&  0.2& 24& 25.4608& $-$1.524$\times$10$^{-1}$& $-$9.230$\times$10$^{-2}$& $-$3.712$\times$10$^{-1}$& $-$2.357$\times$10$^{-1}$&  1.952$\times$10$^{-1}$&    1.661$\times$10$^{-1}$\\
  1.6&  0.2&  0&  0& $-$7.596$\times$10$^{-2}$& $-$7.616$\times$10$^{-2}$& $-$2.171$\times$10$^{-1}$& $-$2.176$\times$10$^{-1}$&  0&    0\\
  1.6&  0.2&  8&  8.6935& $-$8.111$\times$10$^{-2}$& $-$7.641$\times$10$^{-2}$& $-$2.292$\times$10$^{-1}$& $-$2.177$\times$10$^{-1}$&  5.520$\times$10$^{-2}$&    4.502$\times$10$^{-2}$\\
  1.6&  0.2& 16& 17.1994& $-$9.649$\times$10$^{-2}$& $-$7.798$\times$10$^{-2}$& $-$2.647$\times$10$^{-1}$& $-$2.198$\times$10$^{-1}$&  1.032$\times$10$^{-1}$&    8.500$\times$10$^{-2}$\\
  1.6&  0.2& 24& 25.3891& $-$1.221$\times$10$^{-1}$& $-$8.314$\times$10$^{-2}$& $-$3.212$\times$10$^{-1}$& $-$2.288$\times$10$^{-1}$&  1.388$\times$10$^{-1}$&    1.160$\times$10$^{-1}$\\
  1.7&  0.2&  0&  0& $-$6.765$\times$10$^{-2}$& $-$6.799$\times$10$^{-2}$& $-$2.057$\times$10$^{-1}$& $-$2.068$\times$10$^{-1}$&  0&    0\\
  1.7&  0.2&  8&  8.6329& $-$7.116$\times$10$^{-2}$& $-$6.813$\times$10$^{-2}$& $-$2.144$\times$10$^{-1}$& $-$2.066$\times$10$^{-1}$&  3.963$\times$10$^{-2}$&    3.176$\times$10$^{-2}$\\
  1.7&  0.2& 16& 17.1064& $-$8.171$\times$10$^{-2}$& $-$6.995$\times$10$^{-2}$& $-$2.398$\times$10$^{-1}$& $-$2.096$\times$10$^{-1}$&  7.441$\times$10$^{-2}$&    6.024$\times$10$^{-2}$\\
  1.7&  0.2& 24& 25.3085& $-$9.948$\times$10$^{-2}$& $-$7.443$\times$10$^{-2}$& $-$2.806$\times$10$^{-1}$& $-$2.178$\times$10$^{-1}$&  1.009$\times$10$^{-1}$&    8.290$\times$10$^{-2}$\\
  1.7&  0.2& 32& 33.2037& $-$1.257$\times$10$^{-1}$& $-$8.558$\times$10$^{-2}$& $-$3.371$\times$10$^{-1}$& $-$2.366$\times$10$^{-1}$&  1.175$\times$10$^{-1}$&    9.806$\times$10$^{-2}$\\
  1.8&  0.2&  0&  0& $-$5.965$\times$10$^{-2}$& $-$5.962$\times$10$^{-2}$& $-$1.927$\times$10$^{-1}$& $-$1.926$\times$10$^{-1}$&  0&    0\\
  1.8&  0.2&  8&  8.5789& $-$6.211$\times$10$^{-2}$& $-$5.997$\times$10$^{-2}$& $-$1.990$\times$10$^{-1}$& $-$1.930$\times$10$^{-1}$&  2.919$\times$10$^{-2}$&    2.300$\times$10$^{-2}$\\
  1.8&  0.2& 16& 17.0211& $-$6.953$\times$10$^{-2}$& $-$6.147$\times$10$^{-2}$& $-$2.175$\times$10$^{-1}$& $-$1.954$\times$10$^{-1}$&  5.504$\times$10$^{-2}$&    4.380$\times$10$^{-2}$\\
  1.8&  0.2& 24& 25.2283& $-$8.216$\times$10$^{-2}$& $-$6.502$\times$10$^{-2}$& $-$2.474$\times$10$^{-1}$& $-$2.016$\times$10$^{-1}$&  7.515$\times$10$^{-2}$&    6.068$\times$10$^{-2}$\\
  1.8&  0.2& 32& 33.1656& $-$1.009$\times$10$^{-1}$& $-$7.410$\times$10$^{-2}$& $-$2.890$\times$10$^{-1}$& $-$2.190$\times$10$^{-1}$&  8.839$\times$10$^{-2}$&    7.258$\times$10$^{-2}$\\
  1.9&  0.2&  0&  0& $-$5.218$\times$10$^{-2}$& $-$5.210$\times$10$^{-2}$& $-$1.786$\times$10$^{-1}$& $-$1.783$\times$10$^{-1}$&  0&    0\\
  1.9&  0.2&  8&  8.5312& $-$5.394$\times$10$^{-2}$& $-$5.244$\times$10$^{-2}$& $-$1.833$\times$10$^{-1}$& $-$1.787$\times$10$^{-1}$&  2.197$\times$10$^{-2}$&    1.704$\times$10$^{-2}$\\
  1.9&  0.2& 16& 16.9441& $-$5.928$\times$10$^{-2}$& $-$5.373$\times$10$^{-2}$& $-$1.970$\times$10$^{-1}$& $-$1.807$\times$10$^{-1}$&  4.156$\times$10$^{-2}$&    3.254$\times$10$^{-2}$\\
  1.9&  0.2& 24& 25.1518& $-$6.843$\times$10$^{-2}$& $-$5.669$\times$10$^{-2}$& $-$2.192$\times$10$^{-1}$& $-$1.858$\times$10$^{-1}$&  5.706$\times$10$^{-2}$&    4.535$\times$10$^{-2}$\\
  1.9&  0.2& 32& 33.1207& $-$8.213$\times$10$^{-2}$& $-$6.277$\times$10$^{-2}$& $-$2.502$\times$10$^{-1}$& $-$1.966$\times$10$^{-1}$&  6.767$\times$10$^{-2}$&    5.475$\times$10$^{-2}$\\
  2.0&  0.2&  0&  0& $-$4.528$\times$10$^{-2}$& $-$4.530$\times$10$^{-2}$& $-$1.637$\times$10$^{-1}$& $-$1.638$\times$10$^{-1}$&  0&    0\\
  2.0&  0.2&  8&  8.4891& $-$4.657$\times$10$^{-2}$& $-$4.557$\times$10$^{-2}$& $-$1.671$\times$10$^{-1}$& $-$1.641$\times$10$^{-1}$&  1.679$\times$10$^{-2}$&    1.283$\times$10$^{-2}$\\
  2.0&  0.2& 16& 16.8749& $-$5.049$\times$10$^{-2}$& $-$4.664$\times$10$^{-2}$& $-$1.774$\times$10$^{-1}$& $-$1.657$\times$10$^{-1}$&  3.184$\times$10$^{-2}$&    2.457$\times$10$^{-2}$\\
  2.0&  0.2& 24& 25.0802& $-$5.725$\times$10$^{-2}$& $-$4.904$\times$10$^{-2}$& $-$1.941$\times$10$^{-1}$& $-$1.696$\times$10$^{-1}$&  4.391$\times$10$^{-2}$&    3.440$\times$10$^{-2}$\\
  2.0&  0.2& 32& 33.0730& $-$6.743$\times$10$^{-2}$& $-$5.427$\times$10$^{-2}$& $-$2.175$\times$10$^{-1}$& $-$1.793$\times$10$^{-1}$&  5.243$\times$10$^{-2}$&    4.184$\times$10$^{-2}$\\
\hline
 1.5&  0.3&  0&  0& $-$8.481$\times$10$^{-2}$& $-$8.478$\times$10$^{-2}$& $-$2.094$\times$10$^{-1}$& $-$2.094$\times$10$^{-1}$&  0&    0\\
  1.5&  0.3&  8&  8.7037& $-$1.006$\times$10$^{-1}$& $-$7.824$\times$10$^{-2}$& $-$2.442$\times$10$^{-1}$& $-$1.934$\times$10$^{-1}$&  1.484$\times$10$^{-1}$&    1.301$\times$10$^{-1}$\\
  1.5&  0.3& 16& 17.2003& $-$1.469$\times$10$^{-1}$& $-$7.811$\times$10$^{-2}$& $-$3.435$\times$10$^{-1}$& $-$1.864$\times$10$^{-1}$&  2.766$\times$10$^{-1}$&    2.440$\times$10$^{-1}$\\
  1.6&  0.3&  0&  0& $-$8.144$\times$10$^{-2}$& $-$8.123$\times$10$^{-2}$& $-$2.183$\times$10$^{-1}$& $-$2.178$\times$10$^{-1}$&  0&    0\\
  1.6&  0.3&  8&  8.6498& $-$9.182$\times$10$^{-2}$& $-$7.807$\times$10$^{-2}$& $-$2.426$\times$10$^{-1}$& $-$2.095$\times$10$^{-1}$&  1.028$\times$10$^{-1}$&    8.918$\times$10$^{-2}$\\
  1.6&  0.3& 16& 17.1246& $-$1.223$\times$10$^{-1}$& $-$8.089$\times$10$^{-2}$& $-$3.122$\times$10$^{-1}$& $-$2.144$\times$10$^{-1}$&  1.928$\times$10$^{-1}$&    1.683$\times$10$^{-1}$\\
  1.6&  0.3& 24& 25.3046& $-$1.716$\times$10$^{-1}$& $-$8.666$\times$10$^{-2}$& $-$4.197$\times$10$^{-1}$& $-$2.229$\times$10$^{-1}$&  2.607$\times$10$^{-1}$&    2.295$\times$10$^{-1}$\\
  1.7&  0.3&  0&  0& $-$7.362$\times$10$^{-2}$& $-$7.314$\times$10$^{-2}$& $-$2.104$\times$10$^{-1}$& $-$2.095$\times$10$^{-1}$&  0&    0\\
  1.7&  0.3&  8&  8.5953& $-$8.060$\times$10$^{-2}$& $-$7.224$\times$10$^{-2}$& $-$2.277$\times$10$^{-1}$& $-$2.065$\times$10$^{-1}$&  7.240$\times$10$^{-2}$&    6.224$\times$10$^{-2}$\\
  1.7&  0.3& 16& 17.0415& $-$1.013$\times$10$^{-1}$& $-$7.369$\times$10$^{-2}$& $-$2.774$\times$10$^{-1}$& $-$2.084$\times$10$^{-1}$&  1.365$\times$10$^{-1}$&    1.180$\times$10$^{-1}$\\
  1.7&  0.3& 24& 25.2339& $-$1.349$\times$10$^{-1}$& $-$7.800$\times$10$^{-2}$& $-$3.547$\times$10$^{-1}$& $-$2.153$\times$10$^{-1}$&  1.861$\times$10$^{-1}$&    1.622$\times$10$^{-1}$\\
  1.8&  0.3&  0&  0& $-$6.488$\times$10$^{-2}$& $-$6.484$\times$10$^{-2}$& $-$1.973$\times$10$^{-1}$& $-$1.972$\times$10$^{-1}$&  0&   0\\
  1.8&  0.3&  8&  8.5454& $-$6.970$\times$10$^{-2}$& $-$6.480$\times$10$^{-2}$& $-$2.099$\times$10$^{-1}$& $-$1.966$\times$10$^{-1}$&  5.206$\times$10$^{-2}$&    4.436$\times$10$^{-2}$\\
  1.8&  0.3& 16& 16.9628& $-$8.402$\times$10$^{-2}$& $-$6.671$\times$10$^{-2}$& $-$2.461$\times$10$^{-1}$& $-$1.998$\times$10$^{-1}$&  9.857$\times$10$^{-2}$&    8.445$\times$10$^{-2}$\\
  1.8&  0.3& 24& 25.1601& $-$1.075$\times$10$^{-1}$& $-$7.030$\times$10$^{-2}$& $-$3.026$\times$10$^{-1}$& $-$2.056$\times$10$^{-1}$&  1.353$\times$10$^{-1}$&    1.169$\times$10$^{-1}$\\
  1.8&  0.3& 32& 33.1047& $-$1.404$\times$10$^{-1}$& $-$8.153$\times$10$^{-2}$& $-$3.762$\times$10$^{-1}$& $-$2.255$\times$10$^{-1}$&  1.600$\times$10$^{-1}$&    1.394$\times$10$^{-1}$\\
  1.9&  0.3&  0&  0& $-$5.669$\times$10$^{-2}$& $-$5.690$\times$10$^{-2}$& $-$1.829$\times$10$^{-1}$& $-$1.832$\times$10$^{-1}$&  0&    0\\
  1.9&  0.3&  8&  8.5010& $-$6.010$\times$10$^{-2}$& $-$5.683$\times$10$^{-2}$& $-$1.922$\times$10$^{-1}$& $-$1.824$\times$10$^{-1}$&  3.823$\times$10$^{-2}$&    3.229$\times$10$^{-2}$\\
  1.9&  0.3& 16& 16.8911& $-$7.025$\times$10$^{-2}$& $-$5.818$\times$10$^{-2}$& $-$2.189$\times$10$^{-1}$& $-$1.844$\times$10$^{-1}$&  7.263$\times$10$^{-2}$&    6.165$\times$10$^{-2}$\\
  1.9&  0.3& 24& 25.0887& $-$8.701$\times$10$^{-2}$& $-$6.054$\times$10$^{-2}$& $-$2.609$\times$10$^{-1}$& $-$1.874$\times$10$^{-1}$&  1.003$\times$10$^{-1}$&    8.579$\times$10$^{-2}$\\
  1.9&  0.3& 32& 33.0624& $-$1.106$\times$10$^{-1}$& $-$6.912$\times$10$^{-2}$& $-$3.157$\times$10$^{-1}$& $-$2.034$\times$10$^{-1}$&  1.195$\times$10$^{-1}$&    1.032$\times$10$^{-1}$\\
  2.0&  0.3&  0&  0& $-$4.953$\times$10$^{-2}$& $-$4.946$\times$10$^{-2}$& $-$1.683$\times$10$^{-1}$& $-$1.683$\times$10$^{-1}$&  0&    0\\
  2.0&  0.3&  8&  8.4616& $-$5.199$\times$10$^{-2}$& $-$4.970$\times$10$^{-2}$& $-$1.753$\times$10$^{-1}$& $-$1.685$\times$10$^{-1}$&  2.862$\times$10$^{-2}$&    2.395$\times$10$^{-2}$\\
  2.0&  0.3& 16& 16.8262& $-$5.932$\times$10$^{-2}$& $-$5.079$\times$10$^{-2}$& $-$1.954$\times$10$^{-1}$& $-$1.699$\times$10$^{-1}$&  5.452$\times$10$^{-2}$&    4.585$\times$10$^{-2}$\\
  2.0&  0.3& 24& 25.0215& $-$7.150$\times$10$^{-2}$& $-$5.328$\times$10$^{-2}$& $-$2.269$\times$10$^{-1}$& $-$1.737$\times$10$^{-1}$&  7.564$\times$10$^{-2}$&    6.411$\times$10$^{-2}$\\
  2.0&  0.3& 32& 33.0172& $-$8.878$\times$10$^{-2}$& $-$6.003$\times$10$^{-2}$& $-$2.682$\times$10$^{-1}$& $-$1.864$\times$10$^{-1}$&  9.077$\times$10$^{-2}$&    7.771$\times$10$^{-2}$\\
\hline
\end{tabular}
\caption{As in Table~\ref{teuk_ecc1}, but for additional values of
eccentricity $e$; the Teukolsky-based fluxes for $E$ and $L_z$ have an accuracy of $10^{-3}$.\label{teuk_ecc2}}
\end{table}

\begin{table}[h]
\begin{tabular}{|c|c|c|c|c|c|c|c|c|c|}
\hline
 \T\B$\frac{p}{M}$ & $e$ & $\theta_{\rm inc}$& $\iota$& $\frac{dE}{dt}\times\frac{M^2}{\mu^2}$ & $\frac{dE}{dt}\times\frac{M^2}{\mu^2}$ & $\frac{dL_z}{dt}\times\frac{M}{\mu^2}$ & $\frac{dL_z}{dt}\times\frac{M}{\mu^2}$ & $\frac{d\iota}{dt}\times\frac{M}{\mu^2}$& $\frac{d\theta_{\rm inc}}{dt}\times\frac{M}{\mu^2}$ \\
&&(deg.)&(deg.)& (kludge) & (Teukolsky ) & (kludge) & (Teukolsky)& (kludge) &  (kludge) \\
\hline
\T  1.6&  0.4&  0&  0& $-$7.766$\times$10$^{-2}$& $-$7.772$\times$10$^{-2}$& $-$1.918$\times$10$^{-1}$& $-$1.919$\times$10$^{-1}$&  0&    0\\
  1.6&  0.4&  8&  8.5863& $-$9.433$\times$10$^{-2}$& $-$7.645$\times$10$^{-2}$& $-$2.297$\times$10$^{-1}$& $-$1.881$\times$10$^{-1}$&  1.528$\times$10$^{-1}$&    1.370$\times$10$^{-1}$\\
  1.6&  0.4& 16& 17.0151& $-$1.432$\times$10$^{-1}$& $-$7.651$\times$10$^{-2}$& $-$3.382$\times$10$^{-1}$& $-$1.837$\times$10$^{-1}$&  2.873$\times$10$^{-1}$&    2.584$\times$10$^{-1}$\\
  1.7&  0.4&  0&  0& $-$7.882$\times$10$^{-2}$& $-$7.953$\times$10$^{-2}$& $-$2.097$\times$10$^{-1}$& $-$2.115$\times$10$^{-1}$&  0&    0\\
  1.7&  0.4&  8&  8.5426& $-$9.002$\times$10$^{-2}$& $-$7.408$\times$10$^{-2}$& $-$2.367$\times$10$^{-1}$& $-$1.978$\times$10$^{-1}$&  1.087$\times$10$^{-1}$&    9.656$\times$10$^{-2}$\\
  1.7&  0.4& 16& 16.9502& $-$1.229$\times$10$^{-1}$& $-$7.682$\times$10$^{-2}$& $-$3.143$\times$10$^{-1}$& $-$2.025$\times$10$^{-1}$&  2.054$\times$10$^{-1}$&    1.830$\times$10$^{-1}$\\
  1.7&  0.4& 24& 25.1282& $-$1.760$\times$10$^{-1}$& $-$8.090$\times$10$^{-2}$& $-$4.336$\times$10$^{-1}$& $-$2.075$\times$10$^{-1}$&  2.809$\times$10$^{-1}$&    2.514$\times$10$^{-1}$\\
  1.8&  0.4&  0&  0& $-$7.107$\times$10$^{-2}$& $-$7.007$\times$10$^{-2}$& $-$2.013$\times$10$^{-1}$& $-$1.988$\times$10$^{-1}$&  0&    0\\
  1.8&  0.4&  8&  8.4989& $-$7.877$\times$10$^{-2}$& $-$7.001$\times$10$^{-2}$& $-$2.209$\times$10$^{-1}$& $-$1.981$\times$10$^{-1}$&  7.788$\times$10$^{-2}$&    6.879$\times$10$^{-2}$\\
  1.8&  0.4& 16& 16.8817& $-$1.015$\times$10$^{-1}$& $-$7.009$\times$10$^{-2}$& $-$2.774$\times$10$^{-1}$& $-$1.965$\times$10$^{-1}$&  1.478$\times$10$^{-1}$&    1.309$\times$10$^{-1}$\\
  1.8&  0.4& 24& 25.0646& $-$1.383$\times$10$^{-1}$& $-$7.314$\times$10$^{-2}$& $-$3.646$\times$10$^{-1}$& $-$2.003$\times$10$^{-1}$&  2.036$\times$10$^{-1}$&    1.810$\times$10$^{-1}$\\
  1.8&  0.4& 32& 33.0184& $-$1.887$\times$10$^{-1}$& $-$9.193$\times$10$^{-2}$& $-$4.755$\times$10$^{-1}$& $-$2.319$\times$10$^{-1}$&  2.414$\times$10$^{-1}$&    2.156$\times$10$^{-1}$\\
  1.9&  0.4&  0&  0& $-$6.187$\times$10$^{-2}$& $-$6.267$\times$10$^{-2}$& $-$1.861$\times$10$^{-1}$& $-$1.881$\times$10$^{-1}$&  0&    0\\
  1.9&  0.4&  8&  8.4591& $-$6.728$\times$10$^{-2}$& $-$6.216$\times$10$^{-2}$& $-$2.006$\times$10$^{-1}$& $-$1.861$\times$10$^{-1}$&  5.666$\times$10$^{-2}$&    4.980$\times$10$^{-2}$\\
  1.9&  0.4& 16& 16.8173& $-$8.328$\times$10$^{-2}$& $-$6.222$\times$10$^{-2}$& $-$2.424$\times$10$^{-1}$& $-$1.844$\times$10$^{-1}$&  1.079$\times$10$^{-1}$&    9.506$\times$10$^{-2}$\\
  1.9&  0.4& 24& 25.0006& $-$1.094$\times$10$^{-1}$& $-$6.486$\times$10$^{-2}$& $-$3.071$\times$10$^{-1}$& $-$1.878$\times$10$^{-1}$&  1.495$\times$10$^{-1}$&    1.322$\times$10$^{-1}$\\
  1.9&  0.4& 32& 32.9804& $-$1.452$\times$10$^{-1}$& $-$7.884$\times$10$^{-2}$& $-$3.896$\times$10$^{-1}$& $-$2.158$\times$10$^{-1}$&  1.787$\times$10$^{-1}$&    1.588$\times$10$^{-1}$\\
  2.0&  0.4&  0&  0& $-$5.483$\times$10$^{-2}$& $-$5.457$\times$10$^{-2}$& $-$1.735$\times$10$^{-1}$& $-$1.729$\times$10$^{-1}$&  0&    0\\
  2.0&  0.4&  8&  8.4235& $-$5.871$\times$10$^{-2}$& $-$5.445$\times$10$^{-2}$& $-$1.844$\times$10$^{-1}$& $-$1.720$\times$10$^{-1}$&  4.222$\times$10$^{-2}$&    3.686$\times$10$^{-2}$\\
  2.0&  0.4& 16& 16.7586& $-$7.020$\times$10$^{-2}$& $-$5.555$\times$10$^{-2}$& $-$2.158$\times$10$^{-1}$& $-$1.733$\times$10$^{-1}$&  8.064$\times$10$^{-2}$&    7.057$\times$10$^{-2}$\\
  2.0&  0.4& 24& 24.9396& $-$8.902$\times$10$^{-2}$& $-$5.844$\times$10$^{-2}$& $-$2.645$\times$10$^{-1}$& $-$1.778$\times$10$^{-1}$&  1.122$\times$10$^{-1}$&    9.860$\times$10$^{-2}$\\
  2.0&  0.4& 32& 32.9389& $-$1.150$\times$10$^{-1}$& $-$6.536$\times$10$^{-2}$& $-$3.267$\times$10$^{-1}$& $-$1.896$\times$10$^{-1}$&  1.351$\times$10$^{-1}$&    1.193$\times$10$^{-1}$\\
\hline
1.7&  0.5&  0&  0& $-$7.421$\times$10$^{-2}$& $-$7.401$\times$10$^{-2}$& $-$1.815$\times$10$^{-1}$& $-$1.810$\times$10$^{-1}$&  0&    0\\
  1.7&  0.5&  8&  8.4736& $-$8.957$\times$10$^{-2}$& $-$7.168$\times$10$^{-2}$& $-$2.173$\times$10$^{-1}$& $-$1.750$\times$10$^{-1}$&  1.379$\times$10$^{-1}$&    1.256$\times$10$^{-1}$\\
  1.7&  0.5& 16& 16.8300& $-$1.347$\times$10$^{-1}$& $-$6.999$\times$10$^{-2}$& $-$3.201$\times$10$^{-1}$& $-$1.676$\times$10$^{-1}$&  2.611$\times$10$^{-1}$&    2.378$\times$10$^{-1}$\\
  1.8&  0.5&  0&  0& $-$7.589$\times$10$^{-2}$& $-$7.620$\times$10$^{-2}$& $-$1.993$\times$10$^{-1}$& $-$2.000$\times$10$^{-1}$&  0&    0\\
  1.8&  0.5&  8&  8.4395& $-$8.644$\times$10$^{-2}$& $-$6.929$\times$10$^{-2}$& $-$2.254$\times$10$^{-1}$& $-$1.829$\times$10$^{-1}$&  1.005$\times$10$^{-1}$&    9.076$\times$10$^{-2}$\\
  1.8&  0.5& 16& 16.7776& $-$1.175$\times$10$^{-1}$& $-$7.210$\times$10$^{-2}$& $-$3.004$\times$10$^{-1}$& $-$1.880$\times$10$^{-1}$&  1.911$\times$10$^{-1}$&    1.726$\times$10$^{-1}$\\
  1.8&  0.5& 24& 24.9413& $-$1.678$\times$10$^{-1}$& $-$7.395$\times$10$^{-2}$& $-$4.158$\times$10$^{-1}$& $-$1.881$\times$10$^{-1}$&  2.638$\times$10$^{-1}$&    2.385$\times$10$^{-1}$\\
  1.9&  0.5&  0&  0& $-$6.646$\times$10$^{-2}$& $-$6.620$\times$10$^{-2}$& $-$1.855$\times$10$^{-1}$& $-$1.849$\times$10$^{-1}$&  0&    0\\
  1.9&  0.5&  8&  8.4059& $-$7.386$\times$10$^{-2}$& $-$6.320$\times$10$^{-2}$& $-$2.048$\times$10$^{-1}$& $-$1.768$\times$10$^{-1}$&  7.312$\times$10$^{-2}$&    6.579$\times$10$^{-2}$\\
  1.9&  0.5& 16& 16.7233& $-$9.572$\times$10$^{-2}$& $-$6.551$\times$10$^{-2}$& $-$2.603$\times$10$^{-1}$& $-$1.809$\times$10$^{-1}$&  1.395$\times$10$^{-1}$&    1.255$\times$10$^{-1}$\\
  1.9&  0.5& 24& 24.8877& $-$1.312$\times$10$^{-1}$& $-$7.087$\times$10$^{-2}$& $-$3.461$\times$10$^{-1}$& $-$1.909$\times$10$^{-1}$&  1.937$\times$10$^{-1}$&    1.744$\times$10$^{-1}$\\
  1.9&  0.5& 32& 32.8741& $-$1.795$\times$10$^{-1}$& $-$8.247$\times$10$^{-2}$& $-$4.544$\times$10$^{-1}$& $-$2.091$\times$10$^{-1}$&  2.320$\times$10$^{-1}$&    2.092$\times$10$^{-1}$\\
  2.0&  0.5&  0&  0& $-$5.987$\times$10$^{-2}$& $-$5.995$\times$10$^{-2}$& $-$1.761$\times$10$^{-1}$& $-$1.763$\times$10$^{-1}$&  0&    0\\
  2.0&  0.5&  8&  8.3750& $-$6.516$\times$10$^{-2}$& $-$5.918$\times$10$^{-2}$& $-$1.906$\times$10$^{-1}$& $-$1.738$\times$10$^{-1}$&  5.456$\times$10$^{-2}$&    4.882$\times$10$^{-2}$\\
  2.0&  0.5& 16& 16.6725& $-$8.081$\times$10$^{-2}$& $-$5.817$\times$10$^{-2}$& $-$2.324$\times$10$^{-1}$& $-$1.694$\times$10$^{-1}$&  1.044$\times$10$^{-1}$&    9.343$\times$10$^{-2}$\\
  2.0&  0.5& 24& 24.8347& $-$1.063$\times$10$^{-1}$& $-$6.254$\times$10$^{-2}$& $-$2.970$\times$10$^{-1}$& $-$1.776$\times$10$^{-1}$&  1.456$\times$10$^{-1}$&    1.304$\times$10$^{-1}$\\
  2.0&  0.5& 32& 32.8378& $-$1.412$\times$10$^{-1}$& $-$6.993$\times$10$^{-2}$& $-$3.787$\times$10$^{-1}$& $-$1.893$\times$10$^{-1}$&  1.756$\times$10$^{-1}$&    1.576$\times$10$^{-1}$\\
  \hline
 \end{tabular}
 \caption{As in Tables~\ref{teuk_ecc1} and \ref{teuk_ecc2}, but for
   different values of eccentricity $e$; the Teukolsky-based fluxes for $E$ and $L_z$ have an accuracy of $10^{-3}$.\label{teuk_ecc3}}
\end{table}

\begin{table}
\begin{tabular}{|c|c|c|c|c|c|}
\hline
\T $e$ & $\theta_{\rm inc}$ & $\iota$ & $\Delta t/M$ & $\Delta \theta_{\rm inc}$ & $\Delta \iota$\\
  & (deg.) & (deg.) &  & (deg.) & (deg.)\\
\hline
\T  0& 0& 0 & 1.250$\times$10$^{6}$& 0& 0\\
  0& 5& 5.355510& 1.217$\times$10$^{6}$& 1.949$\times$10$^{-1}$& 4.954$\times$10$^{-1}$\\
  0& 10& 10.679331& 1.118$\times$10$^{6}$& 3.468$\times$10$^{-1}$& 8.631$\times$10$^{-1}$\\
  0& 15& 15.943192& 9.574$\times$10$^{5}$& 4.236$\times$10$^{-1}$& 1.019\\
  0& 20& 21.125167& 7.446$\times$10$^{5}$& 4.109$\times$10$^{-1}$& 9.440$\times$10$^{-1}$\\
  0& 25& 26.211779& 4.981$\times$10$^{5}$& 3.158$\times$10$^{-1}$& 6.860$\times$10$^{-1}$\\
  0& 30& 31.199048& 2.528$\times$10$^{5}$& 1.732$\times$10$^{-1}$& 3.527$\times$10$^{-1}$\\
  0& 35& 36.092514& 6.584$\times$10$^4$& 4.636$\times$10$^{-2}$& 8.806$\times$10$^{-2}$\\
\hline
\T  0.1&  0&  0&  1.228$\times$10$^{6}$&  0&  0\\
  0.1&  5&  5.351602&  1.198$\times$10$^{6}$&  4.517$\times$10$^{-1}$&  7.766$\times$10$^{-1}$\\
  0.1& 10& 10.671900&  1.103$\times$10$^{6}$&  6.900$\times$10$^{-1}$&  1.236\\
  0.1& 15& 15.932962&  9.426$\times$10$^{5}$&  7.283$\times$10$^{-1}$&  1.344\\
  0.1& 20& 21.113129&  7.315$\times$10$^{5}$&  6.433$\times$10$^{-1}$&  1.187\\
  0.1& 25& 26.199088&  4.900$\times$10$^{5}$&  4.780$\times$10$^{-1}$&  8.547$\times$10$^{-1}$\\
  0.1& 30& 31.186915&  2.513$\times$10$^{5}$&  2.730$\times$10$^{-1}$&  4.585$\times$10$^{-1}$\\
  0.1& 35& 36.082095&  6.589$\times$10$^4$&  8.385$\times$10$^{-2}$&  1.279$\times$10$^{-1}$\\
  0.2&  0&  0&  1.173$\times$10$^{6}$&  0&  0\\
  0.2&  5&  5.339916&  1.150$\times$10$^{6}$&  1.204&  1.598\\
  0.2& 10& 10.649670&  1.064$\times$10$^{6}$&  1.698&  2.331\\
  0.2& 15& 15.902348&  9.043$\times$10$^{5}$&  1.618&  2.293\\
  0.2& 20& 21.077081&  6.980$\times$10$^{5}$&  1.324&  1.900\\
  0.2& 25& 26.161046&  4.693$\times$10$^{5}$&  9.545$\times$10$^{-1}$&  1.351\\
  0.2& 30& 31.150481&  2.486$\times$10$^{5}$&  5.674$\times$10$^{-1}$&  7.711$\times$10$^{-1}$\\
  0.2& 35& 36.050712&  7.562$\times$10$^4$&  2.070$\times$10$^{-1}$&  2.648$\times$10$^{-1}$\\
  0.3&  0&  0&  1.087$\times$10$^{6}$&  0&  0\\
  0.3&  5&  5.320559&  1.069$\times$10$^{6}$&  2.307&  2.788\\
  0.3& 10& 10.612831&  1.001$\times$10$^{6}$&  3.256&  4.007\\
  0.3& 15& 15.851572&  8.454$\times$10$^{5}$&  2.984&  3.741\\
  0.3& 20& 21.017212&  6.483$\times$10$^{5}$&  2.375&  2.998\\
  0.3& 25& 26.097732&  4.408$\times$10$^{5}$&  1.700&  2.129\\
  0.3& 30& 31.089639&  2.493$\times$10$^{5}$&  1.040&  1.276\\
  0.3& 35& 35.997987&  1.108$\times$10$^{5}$&  4.626$\times$10$^{-1}$&  5.569$\times$10$^{-1}$\\
\hline
\end{tabular}
\caption{Variation in the inclination angles $\iota$ and $\theta_{\rm
  inc}$ as well as time needed to reach the separatrix for several
  inspirals through the nearly horizon-skimming regime.  In all of
  these cases, the binary's mass ratio was fixed to $\mu/M = 10^{-6}$,
  the large black hole's spin was fixed to $a = 0.998M$, and the
  orbits were begun at $p = 1.9M$.  The time interval $\Delta t$ is
  the total accumulated time it takes for the inspiralling body to
  reach the separatrix (at which time it rapidly plunges into the
  black hole).  The angles $\Delta\theta_{\rm inc}$ and $\Delta\iota$
  are the total integrated change in these inclination angles that we
  compute.  For the $e = 0$ cases, inspirals are computed using fits
  to the circular-Teukolsky fluxes of $E$ and $L_z$; for eccentric
  orbits we use the kludge fluxes {\eqref{QdotGG}},
  {\eqref{my_kludge_E}} and {\eqref{my_kludge_L}}.  Notice that
  $\Delta \theta_{\rm inc}$ and $\Delta \iota$ are always positive ---
  the inclination angle always increases during the inspiral through
  the nearly horizon-skimming region.  The magnitude of this increase
  never exceeds a few degrees.\label{delta_inc}}
\end{table}

\end{widetext}

\end{document}